\documentclass{elsart}

\usepackage{amsmath}
\usepackage{amssymb}
\usepackage{hyperref}
\usepackage{url}
\usepackage{graphicx}
\usepackage{psfrag}

\def\be{\begin{equation}}
\def\ee{\end{equation}}
\def\ba{\begin{eqnarray}}
\def\ea{\end{eqnarray}}
\def\bs{\begin{subequations}}
\def\es{\end{subequations}}
\def\bc{\begin{center}}
\def\ec{\end{center}}

\def\g{\gamma}

\def\a{\alpha}
\def\b{\beta}
\def\d{\delta}
\def\G{\Gamma}
\def\O{\Omega}

\def\vs{\varsigma}
\def\s{\sigma}
\def\t{\tau}
\def\l{\lambda}
\def\k{\kappa}

\def\p{\partial}
\def\N{\nabla}

\def\cH{{\mathcal H}}
\def\cS{{\mathcal S}}
\def\cL{{\mathcal L}}
\def\vp{\varphi}
\def\om{\omega}
\def\rmi{{\rm i}}
\def\rme{{\rm e}}

\newcommand{\Eq}[1]{(\ref{#1})}

\begin{document}
\begin{frontmatter}
\title{Ghost conditions for Gauss--Bonnet cosmologies}

\author{Gianluca Calcagni},
\ead{g.calcagni@sussex.ac.uk}
\author{Beatriz de Carlos},
\ead{b.de-carlos@sussex.ac.uk}
\author{Antonio De Felice}
\ead{a.de-felice@sussex.ac.uk}
\address{Department of Physics and Astronomy, University of Sussex, Brighton BN1 9QH, UK}
\date{\today}

\begin{abstract}
We investigate the stability against inhomogeneous perturbations and the appearance of ghost modes in Gauss--Bonnet gravitational theories with a non-minimally coupled scalar field, which can be regarded as either the dilaton or a compactification modulus in the context of string theory. Through cosmological linear perturbations we extract four no-ghost and two sub-luminal constraint equations, written in terms of background quantities, which must be satisfied for consistency. We also argue that, for a general action with quadratic Riemann invariants, homogeneous and inhomogeneous perturbations are, in general, inequivalent, and that attractors in the phase space can have ghosts. These results are then generalized to a two-field configuration. Single-field models as candidates for dark energy are explored numerically and severe bounds on the parameter space of initial conditions are placed. A number of cases proposed in the literature are tested and most of them are found to be unstable or observationally unviable.
\end{abstract}


\end{frontmatter}

%%%%%%%%%%%%%%%%%%%%%%%%%%%%%%%%%%%%%%%%%%%%%%%%%%%%%%%%%%%%%%%%%%%%%%%%%%%%%%%%%%%%%%%%%%%%%%%%%%%%%%%%%%%%%%%%%%%%%%%%%%%%%%%%%%%%%%%%%%%%%%%%%%%%%%%%%%%%%%%%%%%%%%%%%%%%%%%%%%%%%%%%%%%%%%%%%%%%%%%%%%%%%%%%%%%%%%%%%%%%%%%%%%%%%%%%%%%%%%%%

\section{Introduction}

General Relativity (GR) is a very accurate and successful theory of classical gravity which would be desirable to embed in a more fundamental theory. The formulation of a quantum theory of gravity has proven to be a very difficult task and, so far, string theory is the only framework (together with loop quantum gravity) within which gravity can be incorporated at quantum level in a consistent way. Therefore it is worth exploring low-energy aspects of gravity considering, at the same time, leading-order string corrections. This leads us to the study of low-energy actions that contain a Gauss--Bonnet (GB) term, which is a particular combination of quadratic Riemann invariants.

One of the most striking peculiarities of string theory is the fact that all couplings, including those to GB terms, are given dynamically, i.e.\ as vacuum expectation values of fields (called, in general, moduli). This is a feature which makes the study of the evolution of the system of gravity plus moduli fields particularly interesting, as done for a Friedmann--Robertson--Walker (FRW) background in \cite{ART}. In four dimensions, the GB term is topological and does not contribute to the dynamics. Hence it is necessary for it to have a nonconstant, moduli-dependent coupling if we want it to play a nontrivial role, unless one appeals to a Lorentz-violating configuration such as the braneworld.

In this article we deal with the cosmological study of GB gravity and focus on a usually overlooked aspect, namely the possibility of having ghosts and other quantum instabilities in the model. A ghost is, by definition, a field whose kinetic term in the action is unbounded from below (roughly speaking, it has the `wrong sign'), which implies both a macroscopic instability which, if not healed, would lead to a breakdown of the theory, and a violation of unitarity.

It is of utmost importance to realise that the presence and nature of ghost modes for a covariant Lagrangian depends on which classical background the theory is assumed to live in. For instance, in string theory the action is naturally expanded around a Minkowski target space. Ghosts in minimally coupled GB and higher-derivative gravities in \emph{constant curvature} backgrounds were studied in \cite{HOW1,HOW2,NS1,NS2,chi}. Those approaches include both the Minkowski (first explored in \cite{zwi}) and de Sitter (dS) cases, and infer the absence of spin-2 ghosts in actions whose higher-order term is a generic function of the GB combination.

One might be tempted to assume that, since de Sitter is a ghost-free vacuum of the GB theory, any cosmological model with such Lagrangian does not suffer from quantum instabilities. However, the FRW curvature invariant $R$ is not constant and the methods used in the above-mentioned papers cannot be applied any longer. Some works have pioneered this issue by studying cosmological (i.e.\ FRW) perturbations for non-minimally coupled actions \cite{HN00,CHC,HN05}, albeit the ghost problem is not explicitly addressed there.

In analogy to the situation when one considers scalar-tensor theories of gravity, where the absence of spin-2 ghosts require constraints on the vacuum expectation value of the scalar (see \cite{Dic61,mae89}), we have extracted a number of constraints on the field-dependent couplings of a GB model with single scalar field in order to avoid the appearance of instabilities. These are obtained by computing the gravitational perturbations about a FRW background and studying the scalar, vector and tensor contributions separately, as done also in \cite{HN00} and subsequent papers. We also imposed that perturbation modes do not propagate faster than light.

We have then applied the obtained results to different single-field models in literature \cite{ART,NOS,CTS}, whose stability was studied only in phase space, that is, at the classical level (see also \cite{CaN}). The \emph{no-ghost} and \emph{sub-luminal} constraints are here regarded as strong selection rules in the space of parameters of the theory, and we provide an algorithm necessary (but not sufficient) to guarantee quantum stability and obtain a suitable acceleration today.

We have found that, for general initial conditions and during cosmologically relevant periods, ghost modes arise in most of the string-inspired models studied so far.\footnote{A parallel study of modified gravity models can be found in \cite{DHT}. These instabilities do not spare even more complicated cosmological models with an arbitrary function of the GB combination \cite{CENOZ}.} By fine tuning the GB coupling constant $\beta$, however, it is possible to get viable models free of instabilities and reproducing the observed acceleration today. In particular, $\beta$ must be so small that the GB term is actually unobservable at late times.

If one regards the GB action as an effective one, there might be higher-order curvature corrections which are subdominant relative to the leading GB term at late times. Hence, if the latter gives rise to an unstable theory, the higher-order ones will do so as well (the no-ghost conditions would still be dominated by linear and quadratic terms, with no appreciable impact from higher powers in $\beta\ll 1$). Conversely, at high energies or very early time, the effective theory might have received higher-order or nonperturbative modifications which are presently out of control in the sense that, from the study of only the GB term, we cannot know whether they are ghost free or how the universe evolved due to their action. Then one should take this other Lagrangian, whatever it is, and study it from the very beginning as a separate case. The key point is that the GB action below is possibly inadequate, by itself, for describing cosmology.

The presence of attractor solutions in phase space does not guarantee by itself quantum stability, and vice versa. In the context of an action with a general quadratic combination of Riemann invariants, we provide general arguments and several concrete examples showing that stability against homogeneous (i.e.\ classical  in phase-space sense) perturbations is unrelated to that against inhomogeneous (i.e.\ cosmological, quantum) ones.\footnote{
Throughout the paper we refer to cosmological perturbations as `quantum' but in fact they are still (semi)classical, since the perturbed equations do not incorporate loop interactions between particles. However, inhomogeneous perturbations are closer to the quantum picture than homogeneous ones, in the sense that they identify particle modes and their tree-level dynamics. As said above, ghost instabilities in general extend beyond the classical level anyway.}

If probability is not conserved in scattering processes, as when ghosts are involved, wild particle creation can occur in a time interval of cosmological length. For this reason, we define as physically viable those solutions that do not have ghost modes \emph{at any time during the evolution of the universe}. This attitude differs somewhat from other interpretations in literature \cite{CHC}, which we shall briefly discuss in Section \ref{lid}. Although the presence of unstable modes may suggest that the adopted background (FRW) is pathological, whereas the theory \emph{per se} (as covariantly formulated) is not, nonetheless this background is the best known model reproducing natural, observed phenomena on large scales, theoretically consistent with the inflationary scenario. Hence, we will consider solutions to be physically unviable if they lead to ghost modes and/or causality violation on FRW. Therefore, here we do not address the perhaps more abstract problem of the `vacuum structure' of these theories. In our opinion, inconsistencies in FRW are sufficient to discriminate among models.

In \cite{Car03,CJM} an estimate for the decay rate of minimally coupled ghost particles into photons via gravitational interaction was calculated. While theories with a Lorentz invariant cut-off are experimentally forbidden, there remains the possibility of a Lorentz-violating cut-off $\Lambda_{\rm UV}\lesssim 3~$MeV. In our framework, ghost modes are non-minimally coupled to gravity, but this can only tighten the bound of~\cite{CJM} (see that reference for details). Moreover, the characteristic energy scale of the graviton-modulus interaction vertex in modulus-driven cosmologies is that of string theory, hence much higher than $\Lambda_{\rm UV}$. Then each case in which we encounter a ghost mode is automatically excluded. Also, we shall use a very compressed time variable $\tau$ which highlights only features extended through cosmologically long time intervals. The simple fact that one can see the ghosts in $\tau$-plots means that there is no hope for the effect of such instabilities to be unobservable.

The paper is organized as follows. In Section~\ref{setup} we write down the covariant and cosmological equations of motion describing theories with a Gauss--Bonnet term and one non-minimally coupled scalar field. In Section~\ref{sol} we present the general structure of the solutions to the cosmological equations outlined in the previous section, describing some numerical issues associated to these solutions. In Section~\ref{pert} we study the tensor, vector and scalar perturbations about a FRW background and derive four conditions for the absence of quantum instabilities in Gauss--Bonnet cosmologies, plus two more conditions on the speed of propagation of tensor and scalar modes. These conditions are applied to some string motivated models existing in the literature in Section~\ref{models}. The problem of suitable initial conditions, giving stable and experimentally viable solutions, is considered in Section~\ref{pscon}; the theory is here regarded as a model which describes the present acceleration of the universe as measured by supernov\ae\ and large-scale structure observations. The influence of other additional scalar degrees of freedom coupled to the GB combination is discussed in Section~\ref{sec2f}. Section~\ref{disc} is devoted to the discussion of our main results.

%%%%%%%%%%%%%%%%%%%%%%%%%%%%%%%%%%%%%%%%%%%%%%%%%%%%%%%%%%%%%%%%%%%%%%%%%%%%%%%%%%%%%%%%%%%%%%%%%%%%%%%%%%%%%%%%%%%%%%%%%%%%%%%%%%%%%%%%%%%%%%%%%%%%%%%%%%%%%%%%%%%%%%%%%%%%%%%%%%%%%%%%%%%%%%%%%%%%%%%%%%%%%%%%%%%%%%%%%%%%%%%%%%%%%%%%%%%%%%%%

\section{Setup} \label{setup}

We are going to study gravity models with a Gauss--Bonnet term, that we write in the standard form
\be\label{GB}
{\mathcal L}_{\rm GB}\equiv Q-4P+R^2 \,, 
\ee
where the Riemann invariants
\be
P\equiv R_{\mu\nu}R^{\mu\nu},\qquad Q\equiv R_{\mu\nu\l\k}R^{\mu\nu\l\k} \,,
\ee
are built out of the Riemann tensor, Ricci tensor and Ricci scalar which are defined as
\ba
&&R^\l_{~\mu\k\nu} \equiv \p_\k \G^\l_{\mu\nu}-\p_\nu \G^\l_{\mu\k}+\G^\s_{\mu\nu}\G^\l_{\k\s}-\G^\s_{\mu\k}\G^\l_{\nu\s} \,,\\
&&R_{\mu\nu}\equiv R^\l_{~\mu\l\nu}\,,\qquad R\equiv R_{\mu\nu}g^{\mu\nu} \,,
\ea
where $\G^\l_{\mu\nu}$ are the Christoffel symbols. Greek indices run from 0 to 3 and Latin ones over spatial coordinates. The metric signature is $({-}{+}{+}{+})$ and we keep the discussion in four dimensions.

%%%%%%%%%%%%%%%%%%%%%%%%%%%%%%%%%%%%%%%%%%%%%%%%%%%%%%%%%%%%%%%%%%%%%%%%%%%%%%%%%%%%%%%%%%%%%%%%%%%%%%%%%%%%%%%%%%%%%%%%

\subsection{Action and equations of motion}

We assume a gravitational action of the form
\ba
\cS &=&\int d^4x\sqrt{-g}\left\{f_1(\phi)\frac{R}{2\kappa^2}+f_2(\phi)[\cL_{\rm GB}+a_4 \kappa^4(\N_\s\phi\N^\s\phi)^2]\right.\nonumber\\
&&\qquad-\left.\frac12\,\omega(\phi)\N_\mu\phi\N^\mu\phi-V(\phi)+{\sum}_i\cL_{\rho_i}\right\},\label{act}
\ea
where $g$ is the determinant of the metric, $\kappa^2=8\pi G=8\pi/m_P^2$ is the gravitational coupling, $\phi$ is a scalar field with potential $V$ (so that it is a modulus of the theory when $V=0$), $f_i$ and $\omega$ are generic functions of $\phi$, $\cL_{\rm GB}$ is the GB invariant defined in Eq. (\ref{GB}), $a_4$ is a constant, and $\cL_{\rho_i}$ are the Lagrangians for fluids minimally coupled to $\phi$.\footnote{Note that $f_2(\phi)=-\frac12 \alpha'\lambda\xi(\phi)$ in the notation of \cite{CHC}.}

The Einstein equations, $\delta {\mathcal S}/\delta g^{\mu\nu}=0$, read
\be
\Sigma_{\mu\nu}= T_{\mu\nu}\,,\label{eineq}
\ee
where 
\ba
\Sigma_{\mu\nu}&\equiv&\frac{f_1}{\kappa^2}R_{\mu\nu}-\,\omega\N_\mu\phi\N_\nu\phi- \frac{1}{\kappa^2}\N_\mu\N_\nu f_1\nonumber\\
&&+a_4\, \kappa^4f_2\N_\s\phi\N^\s\phi(4\N_\mu\phi\N_\nu\phi-g_{\mu\nu}\N_\s\phi\N^\s\phi)\nonumber\\
&&+g_{\mu\nu}\left(\frac{\Box f_1}{\kappa^2}-\frac{f_1}{2\kappa^2}R+4R\Box f_2-8R^{\s\t}\N_\s\N_\t f_2+\frac12\,\omega\N_\s\phi\N^\s\phi+V\right)\nonumber\\
&&-4R\N_\mu\N_\nu f_2-8R_{\mu\nu}\Box f_2-8R_{(\mu}{}^{\s\t}{}_{\nu)}\N_\s\N_\t f_2+16R_{\s(\mu}\N^\s\N_{\nu)} f_2 \;,
\ea
and
\be
T_{\mu\nu} \equiv -\frac{2}{\sqrt{-g}}\sum_i\frac{\delta(\sqrt{-g}\cL_{\rho_i})}{\delta g^{\mu\nu}} \;. \label{defT}
\ee
Here $\Box \equiv \N^\mu\N_\mu=(-g)^{-1/2}\p^\mu (\sqrt{-g}\,\p_\mu)$. In order to get the first equation, we have used the contracted Bianchi identities and the fact that the variation of the GB term in four dimensions is a total derivative (see, e.g., \cite{dew}). With a general function of the Riemann invariants in arbitrary dimensions one would get more (higher-derivative) terms. Taking the trace of Eq.~(\ref{eineq}) we obtain
\ba
\Sigma& \equiv & \Sigma_\mu{}^\mu=\omega\N_\s\phi\N^\s\phi+4V+(3\Box-R)\frac{f_1}{\kappa^2}+4(R\Box-2R^{\s\t}\N_\s\N_\t)f_2 \nonumber \\
&=& T \equiv  T_\mu{}^\mu \;.
\label{tra}
\ea
The equation of motion for the scalar field is
\ba
&&\omega\Box\phi-V_{,\phi}+\tfrac12\,\omega_{,\phi}\N_\mu\phi\N^\mu\phi+f_{1,\phi}\,\frac{R}{2\kappa^2}+f_{2,\phi}\,[\cL_{\rm GB}+a_4\kappa^4(\N_\s\phi\N^\s\phi)^2]\nonumber\\
&&\qquad-4a_4\kappa^4[(\N_\mu\phi\N^\mu\phi)(\N_\s f_2\N^\s\phi+f_2\Box\phi)+2f_2\N^\mu\phi\N^\nu\phi\N_\mu\N_\nu\phi]=0,\nonumber\\\label{eom}
\ea
where differentiation with respect to the scalar field is indicated with a subscript $_{,\phi}$. When $a_4=0$, only the first line is left.

%%%%%%%%%%%%%%%%%%%%%%%%%%%%%%%%%%%%%%%%%%%%%%%%%%%%%%%%%%%%%%%%%%%%%%%%%%%%%%%%%%%%%%%%%%%%%%%%%%%%%%%%%%%%%%%%%%%%%%%%
%%%%%%%%%%%%%%%%%%%%%%%%%%%%%%%%%%%%%%%%%%%%%%%%%%%%%%%%%%%%%%%%%%%%%%%%%%%%%%%%%%%%%%%%%%%%%%%%%%%%%%%%%%%%%%%%%%%%%%%%

\subsection{Cosmological equations}

The Friedmann--Robertson--Walker line element with flat curvature in synchronous time is $ds^2=-dt^2+a^2(t)\, dx_i dx^i$, where $a(t)$ is the scale factor. The nonvanishing curvature tensor components read
\ba
&&R_{0ii0} = a^2(H^2+\dot H),\qquad R_{ikki} = -a^4H^2,\qquad i\neq k,\\
&&R_{00}   = -3(H^2+\dot H),\qquad  R_{ii} = a^2(3H^2+\dot H),\\
&& R = 6(2H^2+\dot H),
\ea
where $H\equiv\dot a/a$ is the Hubble parameter. When each fluid component is perfect, the energy-momentum tensor becomes
\be
T_{\mu\nu}=\sum_i[(\rho_i+p_i)\,u_\mu u_\nu+p_i\,g_{\mu\nu}],
\ee
where $\rho_i$ and $p_i$ are the energy density and pressure of the fluids, and $u^\mu$ is the unit timelike vector tangent to a fluid element's worldline.

The Friedmann equation $\Sigma_{00}=T_{00}$ is
\be
\label{FReq1}
f_1H^2 +\dot{f}_1H+8\kappa^2\dot f_2H^3-a_4\kappa^6f_2\dot\phi^4=\frac{\kappa^2}{3}\left(\rho_\textsc{tot}+\frac12\,\omega\dot{\phi}^2+V\right),
\ee
where $\rho_\textsc{tot}=\sum_i\rho_i$, while Eq. \Eq{eom} becomes
\ba
&&\omega(\ddot{\phi}+3H\dot{\phi})+V_{,\phi}+\frac12\,\omega_{,\phi}\dot\phi^2-24f_{2,\phi}H^2(H^2+\dot H)-\frac{3}{\kappa^2}\,f_{1,\phi}(2H^2+\dot{H})\nonumber\\
&&\qquad+3a_4\kappa^4\dot\phi^2[4f_2(\ddot\phi+H\dot\phi)+f_{2,\phi}\dot\phi^2]=0.\label{pheom}
\ea
We rescale time as 
\be
t\to \tau\equiv (t-t_0)H_0 \, ,
\nonumber
\ee 
where $H_0=H(t_0)$ is the value of the Hubble parameter today. Then the big bang is at $\tau_i\equiv-t_0 H_0$, $ \tau_i <\tau\leq 0$ until today, and $\tau>0$ in the future. In the following, a subscript 0 denotes quantities evaluated today, dots are derivatives with respect to $\tau$, and $H$ is rescaled so that $H_0=a^{-1}da/d\tau=1$ today. Defining 
\ba
\rho_{c0} & \equiv& 3H_0^2/\kappa^2 \;, \nonumber \\
\beta& \equiv & 8(\kappa H_0)^2\approx 6 \times 10^{-120} \;, \nonumber \\
\O_{i,0}& =& \rho_{i,0}/\rho_{c0} \;, \nonumber \\
\varphi& \equiv & \kappa\phi/\sqrt{3} \;, \nonumber \\
U & \equiv & V/\rho_{c0} \;, \nonumber
\ea
and assuming that only matter and radiation contribute to the energy-momentum tensor, one has that Eqs.~\Eq{FReq1} and \Eq{pheom} become
\ba
&&\beta\dot f_2H^3+f_1H^2 +\dot{f}_1H-\tfrac98\, a_4\beta f_2\dot\vp^4=\rho_m+\rho_r+\tfrac12\,\omega\dot\vp^2+U,\label{freq}\\
&&\omega(\ddot{\vp}+3H\dot{\varphi})+U_{,\vp}+\tfrac12\,\omega_{,\vp}\dot\vp^2-\beta f_{2,\vp}H^2(H^2+\dot H)-f_{1,\vp}(2H^2+\dot{H})\nonumber\\
&&\qquad+\tfrac98\,a_4\beta\dot\vp^2[4f_2(\ddot\vp+H\dot\vp)+f_{2,\vp}\dot\vp^2]=0,\label{pheom2}
\ea
where $\rho_m=\Omega_{m,0}/a^3$ and $\rho_r=\Omega_{r,0}/a^4$. The former is a cubic equation in $H$, which can be rewritten as
\be
y(x) \equiv b_1x^3+b_2x^2+b_3x-\rho=0 \;\;,
\label{Frie}
\ee
where
\bs\ba
&&x \equiv H,\\
&&b_1 = \beta\dot f_2,\qquad b_2 = f_1,\qquad b_3 = \dot{f}_1,\\
&&\rho = \rho_m+\rho_r+\tfrac12\,\omega\dot\varphi^2+U+\tfrac98\,a_4\beta f_2\dot\vp^4 \,.
\ea\es
Finally, the trace equation, Eq.~\Eq{tra}, becomes
\be\label{trace}
\omega\dot\vp^2-4U+\ddot f_1+3H\dot f_1+2(2H^2+\dot H)f_1+\beta[H^2\ddot f_2+(3H^2+2\dot H)H\dot f_2]=\rho_m\,.
\ee

%%%%%%%%%%%%%%%%%%%%%%%%%%%%%%%%%%%%%%%%%%%%%%%%%%%%%%%%%%%%%%%%%%%%%%%%%%%%%%%%%%%%%%%%%%%%%%%%%%%%%%%%%%%%%%%%%%%%%%%%

\subsection{Theoretical models}

In the context of string theory, the field $\vp$ can be interpreted as the massless dilaton field arising in the loop expansion of the low-energy effective action, with 
\bs\ba
f_1 &=& -\omega =\rme^{-\sqrt{3}\,\vp},\qquad U = 0,\\
f_2 &=& \tfrac12\,\lambda\, \rme^{-\sqrt{3}\,\vp},\qquad a_4=-1,\label{tree}
\ea\es
in the so-called `string frame' and at tree level, where $\lambda= 1/4,\,1/8,\,0$ for the bosonic, heterotic and type II string, respectively. The coefficient $a_4$ is fixed in order to recover the three-point scattering amplitude for the graviton~\cite{GS,MT}.

Although we have set $D=4$ from the very beginning, in general string and supergravity actions are defined in $D=10$ (or 11) dimensions, and then compactified down to four dimensions (presently we ignore the braneworld case). Each compactification radius is associated with the vacuum expectation value of a modulus field. In the simplified case of a single modulus (one common characteristic length) and the heterotic string ($\lambda=1/8$) with stabilised dilaton, we can identify $\vp$ with such modulus and, following  \cite{ART}, we have
\bs\ba
\label{modu1}
f_1 &=& 1,\qquad \omega = 3/2,\qquad U = 0,\\
f_2 &=& -\tfrac12\,\xi_0\,\ln [2\rme^{\sqrt{3}\,\vp}\eta^4(\rmi\rme^{\sqrt{3}\,\vp})],\qquad a_4=0,
\label{modu2}
\ea\es
where $\eta$ is the Dedekind function and $\xi_0$ is a constant proportional to the 4D trace anomaly that depends on the number of chiral, vector and spin-$3/2$ massless supermultiplets of the $N=2$ sector of the theory. In general, it can be either positive or negative, but it is positive for theories in which not too many vector bosons are present. At large $|\vp|$, $f_2\sim \xi_0 \cosh(\sqrt{3}\,\vp)$, an approximation which was shown to be very accurate even at small field values \cite{EM,YMO}.

It is therefore reasonable to consider general functions 
\ba
f_1 &=& \gamma_{11}\exp (\gamma_{12} \vp),\\
f_2 &=& \gamma_{21}\exp (\gamma_{22} \vp)+\gamma_{23}\exp (-\gamma_{22} \vp),\label{f2}\\
\omega &=& \gamma_{31}\exp (\gamma_{32} \vp),\\
U &=& \gamma_{41}\exp (\gamma_{42} \vp),\label{expot}
\ea
and constrain the constants $\gamma_{i\!j}$ via the background and perturbed equations for real initial conditions. This will be done in Sections \ref{pert}, \ref{models} and \ref{pscon}. In particular, we shall see in Section \ref{scaper} that the magnitude of the factors $\gamma_{i2}$ in the dilaton exponentials is crucial to determine the quantum stability of the solutions, while for the modulus case it is not particularly important.

The choice of potential, given by Eq.~(\ref{expot}), is motivated by particle physics considerations. If we were to identify $\vp$ with the dilaton or modulus of a string model, then the most likely dynamics for these fields, leaving fluxes aside for the time being, would be nonperturbative effects. Those can come from either instantons or gaugino condensation~\cite{DRSW} and are, in general, exponential in nature.

%%%%%%%%%%%%%%%%%%%%%%%%%%%%%%%%%%%%%%%%%%%%%%%%%%%%%%%%%%%%%%%%%%%%%%%%%%%%%%%%%%%%%%%%%%%%%%%%%%%%%%%%%%%%%%%%%%%%%%%%

\section{Semi-analytic and numerical solutions}\label{sol}

In order to solve the Friedmann equation, Eq.~(\ref{Frie}), we introduce some useful quantities \cite{nic93}, 
\ba
r &\equiv& \frac{4(b_2^2-3b_1b_3)^3}{3^6b_1^4}= \frac{4(f_1^2-3\beta\dot{f}_1\dot f_2)^3}{3^6\beta^4\dot f_2^4},\\
x_* &\equiv& -\frac{b_2}{3b_1} =-\frac{f_1}{3\beta\dot f_2},\\
y_* &\equiv& y(x_*) =\frac{2f_1^3}{27\beta^2\dot f_2^2}-\frac{f_1\dot f_1}{3\beta \dot f_2}-\rho.
\ea
If $f_1$ is constant, then $r$ is positive and one can define $h\equiv \sqrt r$, which is the distance (on the $y$ axis) between the inflexion point and the local extremum. In general, for a cubic with \emph{real} coefficients, one has only one real root when $y^2_*>r$, three real roots (two or three coincident) when $y^2_*=r$,\footnote{We exclude this case a priori, since matter/radiation and the scalar field correspond to independent degrees of freedom. However, we are going to see that $y_*^2\approx r$ approximately.} and three distinct real roots when $y^2_*<r$. In the latter case $h$ is well-defined.

Let us write down $r/y_*^2$ as an expansion in $b_1=\dot f_2\beta\ll 1$, which is equivalent to an expansion in $\beta$ if $\dot f_2$ is not fine tuned. The linear term vanishes identically, leaving
\be
\frac r{y_*^2} = 1+\frac{27}{4f_1^2}\left(\dot f_1^2+4\rho f_1\right)\left(\dot f_2\beta\right)^2+O(\beta^3).\label{hy}
\ee
Therefore the condition to have one single root is 
\be\label{condi}
\dot f_1^2+4\rho f_1<0.
\ee
This is not possible if $f_1>0$, since $\rho>0$ for $-\omega\lesssim O(1)$. If Eq.~\Eq{condi} holds, the root is Cardano's solution
\be\label{sroo}
x_1=x_*+p_++p_-\,,
\ee
where
\be
p^3_\pm=\frac{1}{2b_1}\left(-y_*\pm \sqrt{y_*^2-r}\right).
\ee
In order to understand the behaviour of the solutions, we have to expand them up to $O(\beta)$. To lowest order in $\beta$, Eq.~\Eq{sroo} becomes $x_1\approx 3x_*=O(\beta^{-1})$, which is too large compared with today's value $x\sim O(1)$ and is not compatible with observations. This is true as long as $f_1,\dot f_2\sim O(1)$.

The case $f_1<0$ is problematic in a Minkowski background, where the field $\vp$ acquires a constant vacuum value. In a FRW background, the reality of the solutions requires 
\be
g \equiv \sqrt{1+4\rho\frac{f_1}{\dot f_1^2}}\, \in \mathbb{R}\,, 
\ee
which, during radiation domination, becomes
\be\label{BDREF}
(f_{1,\vp}^2-2|f_1|)\,\dot\vp^2> 4|f_1|\,[U+\Omega_{r,0}(1+z)^4]\,,
\ee
where $z\equiv a^{-1}-1$ is the redshift. This theory looks like a Brans--Dicke model with potential, for which  there are quite strong cosmological bounds \cite{DMT} at big bang nucleosynthesis (BBN). In other words, unless the potential is negative and fine tuned to be comparable to $\Omega_{r,0}z_\textsc{bbn}^4\sim 10^{31}$, the field $\vp$ is subdominant to radiation. If so, Eq.~(\ref{BDREF}) cannot be satisfied in general and at early times $f_1$ cannot be negative. This implies the existence of three real roots, of which we have to choose the positive one consistently with the initial data. As soon as the solution is chosen, the dynamics of the universe is set once and for all.

We conclude that a necessary (but not sufficient) condition to have a viable cosmological solution, compatible with BBN, is\footnote{In fact, one can find out Eq.~\Eq{f1pos} by promoting the number of $e$-foldings $N\equiv \ln a$ to a scalar field, that is, a particle degree of freedom (sometimes this is referred to as Nordstr\"om theory \cite{HE}). Then, its kinetic term has a factor $f_1$ in front of it. Eventually one can find via the same trick that $Q_2>0$, see Eq.~\Eq{TTnoghost2}. However, $N$ interacts with $\vp$ and it is not clear in this approach how to define the energy-momentum tensor of each field.}
\be\label{f1pos}
f_1>0.
\ee
The three distinct real roots of the cubic equation can be found via the Lagrange resolvents method (see, for instance, \cite{wiki}). One solution is given by Eq.~\Eq{sroo}, while the others are
\ba
x_2 &=& x_*+\zeta p_++\zeta^2p_-,\\
x_3 &=& x_*+\zeta^2 p_++\zeta p_-,
\ea
where $\zeta=(-1+\rmi\sqrt{3})/2$ is a primitive third root of 1.
Defining $\epsilon_i\equiv {\rm sgn}(\dot f_i)$ and expanding to order $O(\beta)$ we obtain
\ba
x_2 &=& -\frac{|\dot f_1|}{2f_1}\left(\epsilon_2 g+\epsilon_1\right)+O(\beta),\\
x_3 &=& \frac{|\dot f_1|}{2f_1}\left(\epsilon_2 g-\epsilon_1\right)+O(\beta).
\ea
When $\epsilon_2=+1$, $x_2$ is negative and $x_3$ is positive, whereas when $\epsilon_2=-1$ the vice versa is true. In both cases, the positive solution is the same up to $O(\beta^2)$ and reads
\ba
x       &\approx& x^{(0)}+x^{(1)}\beta\label{x},\\
x^{(0)} &=& \frac{|\dot f_1|}{2f_1}(g-\epsilon_1),\label{xappr0}\\
x^{(1)} &=& -\frac{\dot f_2}{2f_1^3 g}\left[(\dot f_1^2+\rho f_1)\,g-(\dot f_1^2+3\rho f_1)\,\epsilon_1\right].
\ea
When $a_4\neq 0$, $g$ must be expanded at the same level (then $a_4=0$ in $x^{(1)}$ effectively). Note that $x^{(0)}$ is the solution of the quadratic Friedmann equation when $\beta=0$. When $f_1=1$, Eq.~\Eq{xappr0} correctly reproduces the standard Friedmann equation, $x^{(0)2}=\rho$. $x^{(1)}$ is well defined in this case, $x^{(1)} = -\dot f_2\rho/2$. Note that there is no degeneracy in the roots, although they are formally given by the same expression once the sign of $\dot f_2$ is chosen. Therefore the cosmological problem is well posed and consistently resolved.

Near a singularity, when the acceleration $\ddot a$ is large enough, the approximated solutions for $H$ diverge relative to the numerical ones, but we have checked for a number of examples that they agree at later times. In the cases described in Section \ref{pscon}, the agreement is very good for not too high redshifts.

%%%%%%%%%%%%%%%%%%%%%%%%%%%%%%%%%%%%%%%%%%%%%%%%%%%%%%%%%%%%%%%%%%%%%%%%%%%%%%%%%%%%%%%%%%%%%%%%%%%%%%%%%%%%%%%%%%%%%%%%

\subsection{Numerical solutions and initial conditions}\label{numic}

Let us consider the Friedmann equation, Eq.~\Eq{freq}. Taking its time derivative and using Eq.~\Eq{pheom2} we have
\be
A_{11}\,\ddot a+A_{12}\,\ddot\varphi=B_1,
\ee
and, rearranging the equation for the scalar field, we obtain
\be
A_{21}\,\ddot a+A_{22}\,\ddot\vp=B_2\; .
\ee
In the previous two equations the coefficients read
\ba
A_{11}&=& \frac{2}{a}(\b f_{2,\vp}\dot\vp H+f_1),\\
A_{12}&=& \b f_{2,\vp}\,H^2+f_{1,\vp},\\
B_1 &=& (3\b f_{2,\vp}H^2+f_{1,\vp})H\dot\vp-(\b f_{2,\vp\vp}H^2 + f_{1,\vp\vp}+3\om)\,\dot\vp^2+2f_1 H^2\nonumber\\
&&\qquad-(3\rho_m+4\rho_r)-\tfrac92\,a_4\,\beta f_2\dot\vp^4 \;,
\ea
and
\ba
A_{21}&=&-\frac1a\,(\b f_{2,\vp}\,H^2+f_{1,\vp}),\\
A_{22}&=&\om+\tfrac92\,a_4\beta f_2\dot\vp^2,\\
B_2&=& f_{1,\vp}H^2-3\om H\dot\vp-\tfrac12\,\om_{,\vp}\,\dot\vp^2-U_{,\vp}-\tfrac98\,a_4\beta\dot\vp^3(4f_2H+ f_{2,\vp}\dot\vp)\,.
\ea
Note that, in our units, both $a$ and $\varphi$ are dimensionless.

If the $2\times 2$ matrix $A$ with elements $A_{ij}$ is not singular (which is true in general, because the Friedmann equation and the equation for the field are independent), we can solve for $\ddot a$ and $\ddot\vp$. In fact, the determinant $\Delta\equiv \det A$ is
\be
\Delta = A_{11}\,A_{22}-A_{12}\,A_{21}=\frac1a [(2\om+9\,a_4\beta f_2\dot\vp^2)\,(\b f_{2,\vp}\dot\vp H+f_1)+(\b f_{2,\vp}\,H^2+f_{1,\vp})^2].
\ee
The equations can be written as $\ddot X=A^{-1} B$, where $X\equiv (a,\vp)^t$ and $B\equiv(B_1,B_2)^t$. Explicitly,
\ba
\ddot a  &=& \frac{B_1\,A_{22}-B_2\,A_{12}}\Delta\,,\label{acceq}\\
\ddot \vp &=& \frac{B_2\,A_{11}-B_1\,A_{21}}\Delta\,.\label{vppeq}
\ea
As we shall see below, the condition
\be\label{acceq2}
\ddot a>0
\ee
at some time restricts the parameter space, so that not all sets of $\gamma_{i\!j}$'s (and, in particular, $\ddot f_2$) will give rise to an acceleration era. The details of the model will depend on the choice for the reference time, which can be either today (late-time acceleration), near $\tau_i$ (inflation), or both. In the latter case there would be two constraints fixing the solutions, but then one would have to take into account the reheating phase in the early-time dynamics.\footnote{Since the scalar field is assumed to be dynamical today, reheating should procede, for instance, via gravitational particle production.} However, our choice for the time variable is not suitable for analyzing the inflationary phase, the interval between $\tau_i$ and $\tau_\textsc{bbn}$ being too squeezed in our scale. Moreover, it seems difficult to reproduce the correct power spectrum in these theories, at least in the case where a collapsing phase precedes the big bang \cite{TBF,AW}. In general, the presence of a nonvanishing potential or cosmological constant helps in finding accelerating solutions.

The set of initial conditions is $\O_{m,0}\approx 0.25$, $\O_{r,0}\approx 8 \times 10^{-5}$, $a(0)=1$, $\dot a(0)=1$, and $\dot \vp(0)=\dot\vp_0$ as given by the solution of the Friedmann equation, Eq.~(\ref{freq}), evaluated today, $\beta\dot f_2(\vp_0)+f_1(\vp_0)+\dot f_1(\vp_0)=\rho_0$:
\be
\tfrac98\,a_4\beta f_2\dot\vp_0^4+\tfrac12\,\omega_0\dot\vp^2_0-[\beta f_{2,\vp}(\vp_0)+f_{1,\vp}(\vp_0)]\dot\vp_0+\left[\O_{m,0}+\O_{r,0}+U_0-f_1(\vp_0)\right]=0,\label{vpeq}
\ee
which is quadratic in $\dot \vp_0$ for $a_4=0$ (i.e. the compactification modulus). In that case, when $\omega_0>0$ and the last coefficient is dominated by $-f_1<0$, there are typically two branches $\dot\vp_0^-<0$ and $\dot\vp_0^+>0$, roughly corresponding to $\epsilon_2=+1$ ($x_3$ root) and $\epsilon_2=-1$ ($x_2$ root), respectively; in general $\dot f_2$ can change sign during the evolution. In the dilaton case, $\omega_0<0$ and there are four solutions: if $a_4=0$, $\vp_0^-$ is the larger dilaton root, which is positive.

There is only one arbitrary initial condition left and one is free to choose it among our variables. The most attractive and physical choice might be today's acceleration, $\ddot a_0$. However, in order to express $\vp_0$ as a function of $\ddot a_0$ one should invert Eq.~\Eq{acceq}, which can only be done numerically. Therefore we shall choose $\vp(0)=\vp_0$ as the free initial condition. As one of other alternatives, one can fix $\vp_0$ and $\dot\vp_0$ and solve for $U_0$. We are also allowed to consider $\beta$ in the interval $0\lesssim\beta\lesssim 1$ without too strong a fine tuning. For an exponential $f_2$, in fact, one can add a positive constant $\vp_*\sim O(10^2)$ to $\vp$ and define an effective coupling $\beta_{\rm eff}\equiv\beta \rme^{\vp_*}$. From now on and unless specified otherwise, we shall drop the subscript from $\beta_{\rm eff}$.

The string nonperturbative regime occurs for $\vp=0$. To impose $\vp_0\ll 1$ today would both be consistent with general relativity bounds in the case of the dilaton (the effective gravitational coupling being proportional to $\rme^\vp$) and allow an evolution qualitatively different from GR in the case of the modulus with stabilized dilaton. However, we shall argue that this choice does not prevent the rising of instabilities.

Since we want to see what the late-time behaviour is for these theories, the equations are integrated forward in time. It should be noted that a backward integration in time may not be faithful because it might lead to an attractor (which is actually a repeller if integrating from an early time up to now) that can be, in general, not consistent with a standard GR evolution. The numerical solutions are found to be well behaved and numerically stable, in the sense that for each run we plotted $|(\Sigma_{00}-T_{00})/(\Sigma_{00}+T_{00})|$ and $|(\Sigma-T)/(\Sigma+T)|$, and checked that the Einstein equations are satisfied by the numerical solutions at least up to one part to $10^7$.

We conclude this section with a remark on the dilaton. Because of the negative sign of $\omega$, dilaton models are characterized by a contracting phase followed by an expanding one. At the minimum ($H=0$), the scale factor must be smaller than at the nucleosynthesis, $a_{\rm min}<a_\textsc{bbn}\sim 10^{-9}$. This constraint drastically restricts the parameter space. The Friedmann equation at the beginning of the expansion is $\rho=0$. Since $\rho_m+\rho_r\gtrsim 10^{31}$, one should impose, neglecting the term in $a_4\beta$, that $|\omega\dot\vp^2/2+U|\sim|\omega\dot\vp^2|\gtrsim 10^{31}$ at $a_{\rm min}$, a condition hard to achieve. Anyway $\ddot a>0$ in this class of models, and we can rule it out.

%%%%%%%%%%%%%%%%%%%%%%%%%%%%%%%%%%%%%%%%%%%%%%%%%%%%%%%%%%%%%%%%%%%%%%%%%%%%%%%%%%%%%%%%%%%%%%%%%%%%%%%%%%%%%%%%%%%%%%%%%%%%%%%%%%%%%%%%%%%%%%%%%%%%%%%%%%%%%%%%%%%%%%%%%%%%%%%%%%%%%%%%%%%%%%%%%%%%%%%%%%%%%%%%%%%%%%%%%%%%%%%%%%%%%%%%%%%%%%%%

\section{Cosmological perturbations and no-ghost constraints}\label{pert}

One of the key ingredients of this work is the non-minimal coupling between the scalar field and the gravitational sector. Such interaction has been introduced in many different contexts and, probably, scalar-tensor theories represent the simplest example. String models typically couple the modulus field not only with the Ricci scalar, as usual in a scalar-tensor theory, but also with the Gauss--Bonnet term: the hope in allowing for a runaway field is that it may also explain today's acceleration of the universe.

It is well known that, for standard scalar-tensor theories with a coupling of the form $\phi\,R$, the scalar field itself cannot acquire negative expectation values, otherwise the graviton would become a ghost. By expanding the action at second order in the perturbations on a Minkowski background, the kinetic term of the graviton tensor modes is $\phi_0\,h^i{}_j\,\Box h^j{}_i$ for a constant $\phi_0$. Therefore it is required that $\phi_0>0$. In the case of a coupling with the GB term, the expansion about a Minkowski background is trivial in four dimensions. This is because $\phi\,\cL_{\rm GB}$ becomes a total derivative, being $\phi=\phi_0$ in the Minkowski vacuum, and the GB term automatically disappears from the equations of motion.

The same effect is obtained in any other background if the dilaton is stabilised by some mechanism. However, in the presence of a time-dependent dilaton things become more complicated. It should be emphasized that this is exactly the scenario discussed in this paper, where the background is not Minkowski but Friedmann--Robertson--Walker. Typically, in these backgrounds the field $\phi$ acquires nontrivial dynamics and one expects that a coupling with the Gauss--Bonnet term may lead to inconsistencies of the theory, such as a change in the sign of the kinetic terms of some physical degrees of freedom. To see more in detail what we mean by this, we consider the gravitational perturbations about a FRW background, using the standard technique of studying the tensor, vector, and scalar contributions separately~\cite{KS,Muk}.

It is convenient to work in conformal time $\eta\equiv\int dt/a$. The linearly perturbed flat, FRW line element can be decomposed as follows
\ba
ds^2&=&a^2(\eta)\,\{-(1+2\Phi)\,d\eta^2+(2\,B_{,i}-S_i)\,d\eta\,dx^i+[h_{i\!j}+2F_{i,j} \nonumber \\
&&\qquad\qquad+(1-2\Psi)\,\d_{ij}  + 2E_{,ij}]\,dx^i\,dx^j\}\ ,
\ea
where $h_{i\!j}$ is a 2-tensor with spatial indices, $S_i$ and $F_i$ are vectors, and $\Phi$, $\Psi$, $B$, and $E$ are scalars.

%%%%%%%%%%%%%%%%%%%%%%%%%%%%%%%%%%%%%%%%%%%%%%%%%%%%%%%%%%%%%%%%%%%%%%%%%%%%%%%%%%%%%%%%%%%%%%%%%%%%%%%%%%%%%%%%%%%%%%%%

\subsection{Tensor perturbations}

The FRW line element at zero curvature is conformally flat and it is possible to choose the transverse-traceless harmonic gauge for the tensor perturbations~\cite{MTW}
\be
h_{\mu 0}=0=h_\mu{}^\mu,\qquad \p_jh_i{}^j{}=0.
\ee
Ignoring the other modes, it is long but not difficult to find that the action at second order in $h_{i\!j}$ is
\begin{eqnarray}
\delta^{(2)} \cS_{\rm TT}&=&\int d\eta\,d^3x\,a^4\left\{\left[\frac{f_1}{8\kappa^2}+\frac1{a^2}\,(f_2''-\cH \,f_2')\right]\,h^j{}_i\Box h^i{}_j \right. \nonumber \\
&&\qquad+\left. \frac1{a^2}\,(f_2''-2\,\cH \,f_2')\,h^j{}_i {h^i{}_j}\!\!'' \right\}+\dots ,
\end{eqnarray}
where $\Box=-\p_\eta^2+\p_x^2$ is the Minkowskian d'Alembertian, primes denote derivatives with respect to $\eta$, $\cH\equiv a'/a$, and we have kept only terms containing second derivatives of the perturbation $h^i{}_j$ (the dots stand for linear-derivative terms, no effective potential). The integrand can be written as 
\begin{equation}
-[K_1(\eta)-K_2(\eta)]\,h^j{}_i\p^2_\eta h^i{}_{j}+K_1(\eta)\,h^j{}_i\p_x^2 h^i{}_{j}\ ,
\end{equation}
where
\begin{eqnarray}
K_1(\eta)&=&\frac{f_1}{8\kappa^2}+\frac1{a^2}\,(f_2'' -\cH \,f_2'),\\
K_2(\eta)&=&\frac1{a^2}\,(f_2''-2\,\cH \,f_2')\ .
\end{eqnarray}
By performing a change of the time variable $\eta\to\bar\eta$, we have that
\begin{equation}
\p^2_\eta h_{ij}=\left(\frac{d\bar\eta}{d\eta}\right)^{\!2} \p^2_{\bar\eta} h_{ij}+\p_{\bar\eta} h_{ij}\,\frac{d^2\bar\eta}{d\eta^2}\ .
\end{equation}
Therefore if
\begin{equation}
\frac{d\bar\eta}{d\eta}=\sqrt{\frac{K_1}{K_1-K_2}}\equiv P^{-1}\ ,
\label{ghost1}
\end{equation}
the second-derivative operator becomes 
\begin{equation}
\int d\bar\eta\, d^3x\,a^4\, P(\bar\eta)\, K_1(\bar\eta)\,h^j{}_i\,\Box h^i{}_j+\dots\,.
\label{ghost2}
\end{equation}
We require the theory to be ghost free. Combining~(\ref{ghost1}) and~(\ref{ghost2}), one has a ghost if
$K_1<0$ and, because of Eq.~\Eq{ghost1}, for $K_1-K_2<0$. In other words the kinetic term of the field is well defined only if
\begin{equation}
K_1>0\qquad{\rm and}\qquad K_1-K_2>0\ .
\end{equation}
In universal time the above conditions can be written as
\begin{eqnarray}
f_1+8\,\kappa^2\,\ddot f_2&>&0\ ,\\
f_1+8\,\kappa^2\,H\,\dot f_2&>&0\ ,
\end{eqnarray}
or, in terms of $\tau$ and rescaled units,
\begin{eqnarray}
{\rm TT1:}&\qquad& Q_1\equiv f_1+\beta\,\ddot f_2>0\label{TTnoghost1}\,,\\
{\rm TT2:}&\qquad& Q_2\equiv f_1+\beta\,H\,\dot f_2>0\ .\label{TTnoghost2}
\end{eqnarray}
Incidentally, the speed of propagation $s_\textsc{tt}$ of tensor perturbations is indeed equal to $Q_1/Q_2$. This speed needs to be positive definite, otherwise the system becomes unstable: the second-order differential operator in Eq.~\Eq{ghost2} would become (plus or minus, depending on ${\rm sgn}(K_1)$) a 4D Laplacian rather than a d'Alembertian and perturbations would grow exponentially. Therefore this kind of instability occurs if either TT1 or TT2 are violated, whereas a ghost appears when both these conditions do not hold at the same time.

It is important to notice that the conditions TT1 and TT2 still hold in the presence of more than one scalar field, because the scalar degrees of freedom do not mix with the tensor ones: for example, $f_2$ may be a function of, say, two fields which can be thought of as representing the real and imaginary parts of a complex scalar (see Section \ref{sec2f}). These conditions should hold at all times and a numerical simulation is necessary to make sure that both Eqs.~(\ref{TTnoghost1}) and \Eq{TTnoghost2} are satisfied during the evolution of the universe. It is also clear that, if $\beta$ is small and $f_2$ and its derivatives are always $\lesssim O(1)$, then the equations are satisfied providing that $f_1>0$, which is the no-ghost condition for a scalar-tensor theory deduced already from the background equations, Eq.~\Eq{f1pos}. Of course, for a de Sitter background with constant $\vp$ both conditions hold but, in general, this is not the case (in particular, a dS background is actually achieved also when $\vp\propto \tau$).

We can already restrict the parameter space for the initial conditions by requiring that both TT1 and TT2 should be valid. As an example, for the dilaton, TT2 implies
\be
1-\frac{\lambda}{2}\,\beta\,\dot\vp_0>0\ .
\ee
For values of $\beta\sim 10^{-2}$, TT2 is not satisfied, giving rise to an instability when, for example, $\vp_0\sim 10$ and we choose the branch $\dot\vp_0^+\sim O(10^2)$ ($U=0$ is assumed). The level of fine tuning in this example is still modest, but it is clear that it corresponds to an unphysical situation, since the effective gravitational coupling would be strongly enhanced today.

It is interesting to note that the dilaton case \emph{can} in principle have ghost modes on FRW, although this requires a tuning of the parameters in contrast with experiments. In fact, in these theories, $\dot\varphi_0$ is typically related with the time derivative of the Newton constant, that is $|\gamma_{12}\dot\vp_0|\sim |H_0^{-1}\,G^{-1}dG/dt|_0<1.75\times10^{-2}$ (see~\cite{WTB}). Then it is true that viable string-inspired cosmologies, if any, are ghost-free like the mother theory, but \emph{this property is not automatically satisfied by construction}. This result is independent of the value of $a_4$.

%%%%%%%%%%%%%%%%%%%%%%%%%%%%%%%%%%%%%%%%%%%%%%%%%%%%%%%%%%%%%%%%%%%%%%%%%%%%%%%%%%%%%%%%%%%%%%%%%%%%%%%%%%%%%%%%%%%%%%%%

\subsection{Vector perturbations}

It can be proved that, in the absence of an anisotropic fluid and introducing the quantity $V_i\equiv S_i+a\,\dot F_i$, the following quantity is conserved
\be
a^2\,(f_1+8\,\kappa^2\,H\,\dot f_2)\,V_{(i,j)}={\rm const.}
\ee
This is equivalent to saying that vector perturbations do not propagate like waves or, in other words, the angular momentum of the fluid is conserved (see~\cite{HN00,HN05}). The other equation for the vector perturbation in vacuum states that $V_i$ is an harmonic function satisfying the Laplace equation, $\p_j\p^j V_i=0$.

%%%%%%%%%%%%%%%%%%%%%%%%%%%%%%%%%%%%%%%%%%%%%%%%%%%%%%%%%%%%%%%%%%%%%%%%%%%%%%%%%%%%%%%%%%%%%%%%%%%%%%%%%%%%%%%%%%%%%%%%

\subsection{Scalar perturbations}\label{scaper}

If we can neglect the matter contribution, i.e.~at late times or in vacuum, the calculations simplify considerably by choosing the uniform field gauge $\d\phi=0$. In this case it is possible to show that the action for the potential $\Psi$ can be written as \cite{HN00}
\be
\d^{(2)}\cS_{\rm SC}=\frac12\int d^3x\,dt\,a^3\,Q_{\rm SC}\left[\dot\Psi^2-\frac 
{s_\textsc{sc}}{a^2}\,(\p_i\Psi)^2\right] ,
\ee
where
\be
Q_{\rm SC}\equiv \frac{\om\dot\phi^2+\frac3{2\kappa^2}\frac{(\dot f_1+8\kappa^2\,H^2\,\dot f_2)^2}{f_1+8\kappa^2\,H\,\dot f_2}+12a_4\kappa^4f_2\dot\phi^4}
{\left(H+\tfrac12\frac{\dot f_1+8\kappa^2\,H^2\,\dot f_2}{f_1+8\kappa^2\,H\,\dot f_2}\right)^{\!2}}\ ,
\ee
and
\be
s_\textsc{sc}\equiv 1-4\,\frac{%
\dot f_2\left(\frac{\dot f_1+8\kappa^2\,H^2\,\dot f_2}{f_1+8\kappa^2\,H\,\dot 
f_2}\right)^{\!2}
\left(\frac{\ddot f_2}{\dot f_2}-H-4\dot H\,
\frac{f_1+8\kappa^2\,H\,\dot f_2}{\dot f_1+8\kappa^2\,H^2\,\dot 
f_2}\right)+2a_4\kappa^4f_2\dot\phi^4}%
{\om\dot\phi^2+\frac3{2\kappa^2}\frac{(\dot f_1+8\kappa^2\,H^2\,\dot 
f_2)^2}%
{f_1+8\kappa^2\,H\,\dot f_2}+12a_4\kappa^4f_2\dot\phi^4}\ .
\ee
Therefore, it is clear what the no-ghost conditions for the scalar 
modes are,
\be
Q_{\rm SC}>0\qquad{\rm and}\qquad s_\textsc{sc}>0\ .
\ee
The first condition can be rewritten as
\be\label{scQ}
\kappa^2\om\,\dot\phi^2+\frac32\,\frac{(\dot f_1+8\kappa^2\,H^2\,\dot f_2)^2}%
{f_1+8\kappa^2\,H\,\dot f_2}+12a_4\kappa^6f_2\dot\phi^4>0\ ,
\ee
or
\be
{\rm SC1:}\qquad q\equiv\om\,\dot\vp^2+\frac{Q_3^2}%
{2Q_2}+\frac{27}{2}\,a_4\beta f_2\dot\vp^4>0\,,\label{sc1}
\ee
where
\be
Q_3 \equiv \dot f_1+\beta H^2 \dot f_2=H(Q_2-f_1)+\dot f_1.\label{rel}
\ee
If $\om\geq 0$ and $a_4=0$, this condition does not add anything new to the tensor constraints. 

The second condition, that is trivially satisfied if $\dot f_2=0$ and $a_4=0$, is
\ba
{\rm SC2:}\qquad&& s_\textsc{sc}=1-\frac{\beta}{6q}\left[\dot f_2\left(\dfrac{Q_3}%
{Q_2}\right)^{\!2}%
\left(\dfrac{\ddot f_2}{\dot f_2}-H-4\,\dot H\,
\dfrac{Q_2}{Q_3}\right)\right.\nonumber\\
&&\qquad\qquad\qquad\left.\vphantom{\frac11}+18\,a_4 f_2\dot\vp^4\right]>0\,.\label{sc2}
\ea
This is a new constraint, being independent from the others we have already encountered. Therefore there are, in general, four different \emph {no-ghost conditions} that should be satisfied in order for the theory to be consistent and free of instabilities.\footnote{Strictly speaking, not all violations of these conditions will lead to \emph{ghost} instabilities, as explained below Eq.~\Eq{TTnoghost2}. For convenience, we shall keep calling the above conditions `no-ghost'.} Note that $s_\textsc{sc}$ is interpreted as the speed of propagation of the perturbation.

The no-ghost conditions do not depend on the scalar potential. This is because, intuitively, the sign of $U$ determines whether $\vp$ is a tachyon (which is an instability quite recurrent in FRW), but does not affect the kinetic term. In any case, as we shall see below, the potential does modify the background solution and, indirectly through the no-ghost constraints, their stability.

When $f_2$ is constant or $\beta$ is very small, Eq.~\Eq{sc1} gives a bound on the value of the coefficients $\g_{i\!j}$ which does not depend on $\dot\vp$. Assuming $\g_{32}=\g_{12}$ and $\g_{11}$ as in the dilaton case, one can factorize the exponentials and get ($\beta=0$)
\be\label{brd}
2\frac{\g_{31}}{\g_{11}}+\g_{12}^2>0,
\ee
where we have imposed the TT condition $\g_{11}>0$. This equation is valid in the GR limit, which must be reached at some point during the evolution of the universe. For this reason, actually, it holds at all times as a necessary (but not sufficient) condition. For a minimally coupled scalar field ($\g_{12}=0$) and positive gravitational coupling, this condition is trivially $\g_{31}>0$, while for a dilaton-like field $\g_{11}=-\g_{31}=1$, it implies that $|\g_{12}|>\sqrt{2}$. The string dilaton respects this bound.

Defining 
\be
\phi_\textsc{bd} \equiv \frac{\g_{11}}2\,\rme^{\g_{12}\vp}\,,\qquad \om_\textsc{bd} \equiv \frac{3}{\g_{12}^2}\frac{\g_{31}}{\g_{11}}\,,
\ee
the Lagrangian with $\beta=0$ reduces to a Brans--Dicke theory in Einstein gravity,
\be
\cL_\textsc{bd}=\phi_\textsc{bd} R -\frac{\om_\textsc{bd}}{\phi_\textsc{bd}}\N_\mu\phi_\textsc{bd}\N^\mu\phi_\textsc{bd}-V(\phi_\textsc{bd}),
\ee
and Eq.~\Eq{brd} becomes the usual no-ghost constraint $\om_\textsc{bd}>-3/2$ (together with the TT condition $\phi_\textsc{bd}>0$) for Brans--Dicke theories. 

Note that all the results in the scalar sector are obtained in a gauge-invariant framework \cite{HN05}.\footnote{Actually, in the more general case where the term $f_1(\phi)R$ in the action is replaced by $f(\phi,R)$ for some smooth function $f$, one can find \emph{two} scalar degrees of freedom obeying independent Mukhanov equations. This can be understood by recalling that $f(R)$ gravity has a hidden scalar degree of freedom, which adds to that coming from the GB term. However, one can see that there is always a unique set of no-ghost conditions as follows. 

The two scalar modes are governed by Eqs. (104) and (105) of \cite{HN05} (see the original paper for the notation; here $c_2=c_3=0$), and the former can be written as $(a^3 Q_{\rm SC})^{-1} d(a^3 Q_{\rm SC} \dot\Phi)/dt=s_\textsc{sc} (\p_i\p^i\Phi)/a^2$. Equation (105) can be recast as $(a^3 Q_\Psi)^{-1} d(a^3 Q_\Psi \dot{\tilde\Psi})/dt=s_\textsc{sc}(\p_i\p^i\tilde\Psi)/a^2$, where $\tilde\Psi = \Psi aQ_2/[H+Q_3/(2Q_2)]$ and $Q_\Psi = a^{-4} (Q_{\rm SC} s_\textsc{sc})^{-1}$. After requiring $s_\textsc{sc}>0$, to impose either $Q_\Psi>0$ or $Q_{\rm SC}>0$ results in precisely the same condition.

In the case of study (no anisotropic stress, Einstein--Hilbert leading term) there is only one independent scalar field
which can be canonically quantized, provided the speed of propagation is real. Consistently, in the uniform field gauge ($\delta\phi=0$) $\Phi=\Psi$.}

%%%%%%%%%%%%%%%%%%%%%%%%%%%%%%%%%%%%%%%%%%%%%%%%%%%%%%%%%%%%%%%%%%%%%%%%%%%%%%%%%%%%%%%%%%%%%%%%%%%%%%%%%%%%%%%%%%%%%%%%

\subsection{On superluminal propagation of perturbations}

There is another issue, related to the speed of propagation of the perturbations, which can further constrain cosmological solutions. Particles that propagate at speed faster than light can generate paradoxical situations in which Lorentz invariance is broken and preferential frames in which the physics is well defined are undemocratically selected. In Minkowski background, superluminal propagation is associated with breaking of causality, ill-posed Cauchy problem \cite{AL}, the possibility of closed time-like curves, and violation of the null energy condition. The latter is usually responsible for instabilities, although there are stable systems in which the null energy condition does not hold \cite{DGNR}. Superluminal particles are also associated to the spoiling of conventional black hole dynamics and thermodynamics \cite{DS,BMV}.

To extend this analysis to curved backgrounds is not easy and, to the best of our knowledge, only partial evidence in support of bad causal behaviour has been provided so far~\cite{VBL,AADNR,DS}. For instance, the presence of superluminal modes would imply that for any comoving observer on a inertial reference frame (say, an experiment to detect gravitational waves), the passage of a superluminal signal would be felt as a breakdown of causality. On the other hand, the Mukhanov equations for the perturbations are defined on a conformally flat background, and this alone might be sufficient to justify the imposition of the \emph{sub-luminal} (SL) \emph{constraints}
\ba
{\rm SL1:}\qquad&& s_\textsc{tt}\equiv Q_1/Q_2\leq 1\qquad\Rightarrow\qquad \ddot f_2-H\,\dot f_2\leq 0,\label{slc1}\\
{\rm SL2:}\qquad&& s_\textsc{sc}\leq 1.\label{slc2}
\ea
Our perspective is to dismiss as unviable solutions on which perturbations propagate faster than light. 

Note that, when $a_4=0$, Eq.~\Eq{sc2} can be rewritten as
\be
6q\frac{Q_2}{Q_3^2}(1-s_\textsc{sc}) = -(1-s_\textsc{tt})+4\epsilon \left(1-\frac{\dot f_1}{Q_3}\right),
\ee
where $\epsilon\equiv -\dot H/H^2$. In a non-superaccelerating universe ($\epsilon>0$) in which the SL1 and TT2 conditions are satisfied, the first term is nonpositive, while the second one is non-negative if, and only if,
\be\label{eqsl2}
Q_3 \geq \dot f_1\qquad \Leftrightarrow\qquad \dot f_2\geq 0\,.
\ee
Equation \Eq{eqsl2} is necessary to guarantee the sub-luminal condition SL2, $1-s_\textsc{sc}\geq 0$.

%%%%%%%%%%%%%%%%%%%%%%%%%%%%%%%%%%%%%%%%%%%%%%%%%%%%%%%%%%%%%%%%%%%%%%%%%%%%%%%%%%%%%%%%%%%%%%%%%%%%%%%%%%%%%%%%%%%%%%%%

\subsection{When other fluids are present}

All the above results were found in the absence of other perfect fluids in the action, and it is natural to ask what happens in more realistic situations when both pressureless matter and radiation evolve together with $\phi$.

At late times both matter and radiation are subdominant relative to $\phi$ (regarded as dark energy) and they can be neglected in an asymptotic solution. Therefore, if a solution does not satisfy the ghost constraints at late times, it is sufficient for us to discard it as unviable altogether.

Problems may arise when trying to check a solution which is ghost-free asymptotically in the future, or when one assumes that dark energy is not the modulus but another fluid with $w\approx -1$ today. However, we do not expect the four ghost conditions to be modified in the presence of extra perfect fluids, provided these are \emph{minimally coupled} to both gravity and the scalar field. The tensor constraints would be unchanged since there are no tensor modes in isotropic fluid perturbations. On the other hand, to any new fluid component would correspond another scalar mode to disentangle from that of the modulus perturbation (for a multi-field example, see \cite{GWBM}). 

There is evidence that the constraint equations already written would not be affected. A way to see this is to note that the usual no-ghost constraint $\om_\textsc{bd}>-3/2$, Eq.~\Eq{scQ} for Brans--Dicke theories in Einstein gravity,  comes also from a conformal transformation from the Jordan to the Einstein frame, where the (no-)ghost mode becomes apparent. This constraint does not depend on the frame choice, nor upon the presence of non-minimally coupled perfect fluids. In the moduli case, if Eq.~\Eq{scQ} were modified when a new fluid component is taken into account, the only way to recover the Brans--Dicke case would be to have couplings like $b(f_2,\dot f_2)\,c(\rho,p)$ for some functions $b$ and $c$. But at the level of the background action such couplings are absent, and in the ghost conditions only couplings already present from the beginning can appear (for instance, both $f_1$ and $f_2$ couple to gravity, while $\om$ does not). A direct inspection of the perturbed Einstein equations is difficult for assessing this result more rigorously and, for the moment, we shall regard it as a conjecture yet to be proved.

%%%%%%%%%%%%%%%%%%%%%%%%%%%%%%%%%%%%%%%%%%%%%%%%%%%%%%%%%%%%%%%%%%%%%%%%%%%%%%%%%%%%%%%%%%%%%%%%%%%%%%%%%%%%%%%%%%%%%%%%
%%%%%%%%%%%%%%%%%%%%%%%%%%%%%%%%%%%%%%%%%%%%%%%%%%%%%%%%%%%%%%%%%%%%%%%%%%%%%%%%%%%%%%%%%%%%%%%%%%%%%%%%%%%%%%%%%%%%%%%%

\section{Ghost conditions for models in literature}\label{models}

Although we have considered only inhomogeneous (i.e.\ cosmological) perturbations, it is common in literature to study the attractor solutions of modified gravity models via a phase-space analysis. A natural question one may ask is: can an attractor, stable against homogeneous perturbations, be unstable against cosmological perturbations? The answer is yes, and we shall provide several examples. The inequivalence of the two approaches was stressed also in \cite{CJM,FN}. A weaker (and maybe more obvious) result holds, namely, phase-space unstable solutions can be free of ghosts. 

We can compare the prediction of the no-ghost constraints with some known cosmological expanding solutions \cite{CTS}, with Gauss--Bonnet parametrization in four dimensions. To make contact with past notation, we switch back to synchronous time and set $\kappa^2=1$. 

%%%%%%%%%%%%%%%%%%%%%%%%%%%%%%%%%%%%%%%%%%%%%%%%%%%%%%%%%%%%%%%%%%%%%%%%%%%%%%%%%%%%%%%%%%%%%%%%%%%%%%%%%%%%%%%%%%%%%%%%

\subsection{Asymptotic solutions and modulus scenarios} 

At late times one can identify three different limits: a `low curvature' regime in which the Einstein--Hilbert action dominates over the Gauss--Bonnet term, a `high curvature' regime where the quadratic term dominates over the linear one, and an `exact' regime in which both terms contribute at the same level. Asymptotically, one can consider a power-law expansion, $a\sim t^{\s_1}$, with logarithmic modulus ($\dot f_1=0$)
\be\label{logas}
H\sim\frac{\s_1}t,\qquad \dot\phi=\frac{\s_2}t,\qquad f_2\sim \frac{\xi_0}8\,t^{\s_2},
\ee
where $\s_i$ and $\xi_0$ are determined by which of the three limits above one imposes to reach. Any fluid content has been damped away at earlier times.\footnote{In the notation of \cite{CTS}, $\s_i=\om_i$, $\xi_0=4\tilde\d$. Also, we do not consider solutions with sudden future singularities.} For the compactification modulus ($f_1=1$, $\omega>0$), the no-ghost constraints read
\bs\label{gcasy}\ba
{\rm TT1:}&\qquad& 1+\xi_0\s_2(\s_2-1)\,t^{\s_2-2}>0,\\
{\rm TT2:}&\qquad& 1+\xi_0\s_1\s_2\,t^{\s_2-2}>0,\\
{\rm SC2:}&\qquad& \frac{\s_2^2}{t^2}+\frac{(\xi_0\s_1^2\s_2\,t^{\s_2-3})^2}{1+\xi_0\s_1\s_2\,t^{\s_2-2}} -\frac{(\xi_0\s_2)^3\s_1^4}{3(1+\xi_0\s_1\s_2t^{\s_2-2})^2} \nonumber \\
& \qquad &\qquad \times\left[(\s_2-\s_1+3)t^{3\s_2-8}+\frac{4t^{2(\s_2-3)}}{\xi_0\s_1\s_2}\right]>0. 
\ea\es
Here we do not discuss the viability of these solutions as models of dark energy, and limit the comparison to the stability issue. After looking at attractor solutions, we shall see a couple of numerical examples. In this section only, a solution is said to be stable or unstable in the phase-space sense.

The low-curvature solutions are $\s_1=1/3$, $\s_2^\pm=\pm 2/3$, and both branches are stable.\footnote{If one allows for an extra perfect fluid, it must be stiff matter, $p=\rho$.} The TT and SC conditions are trivial: as regards SC2, the leading term is $\s_2^2t^{-2}>0$.

There is also a high-curvature asymptotic solution for the modulus with zero potential, which requires a nonvanishing cosmological constant \cite{CTS}: $\s_1=1$, $\s_2=4$, $V=12\xi_0$. Such solution is an attractor only when $\xi_0>0$. Both TT conditions require $\xi_0>0$. However, SC2 gives
\be
s_\textsc{sc}=-4\xi_0>0 \qquad\Leftrightarrow\qquad \xi_0<0.
\ee
Therefore the attractor solution has always ghosts. Conversely, non-attractor solutions ($\xi_0<0$) are always ghost-free. According to our interpretation of the ghost conditions, one can infer that it is never possible to achieve a physical high-curvature regime asymptotically.

\begin{figure}
\begin{center}
{\psfrag{t}{$\tau$}
\psfrag{Ht}[][][.8][0]{$\dfrac1\epsilon$}
\includegraphics[width=8cm]{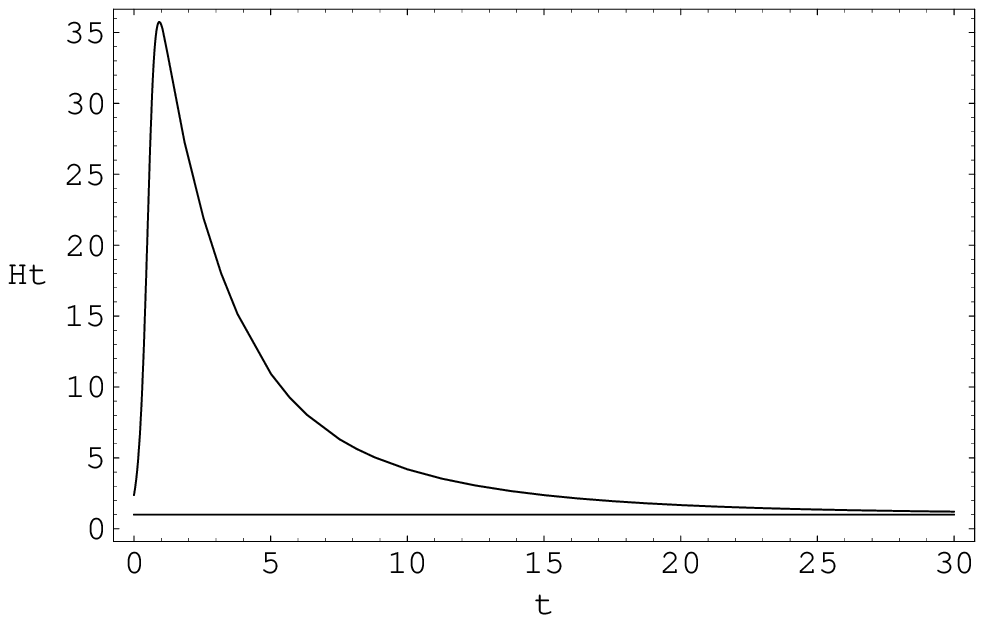}}\\
{\psfrag{t}{$\tau$}
\psfrag{ut}[c][][.8][0]{$\dfrac{\dot\phi}{\epsilon H}\;\;$}
\includegraphics[width=8truecm]{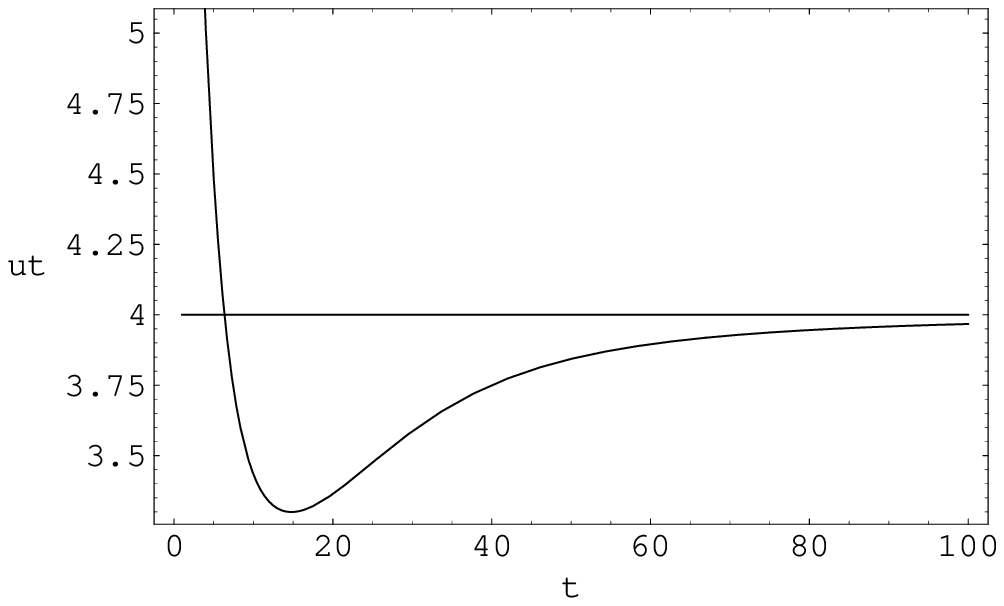}}\\
{\psfrag{t}{$\tau$}
\psfrag{s}{$s_\textsc{sc}$}
\includegraphics[width=8truecm]{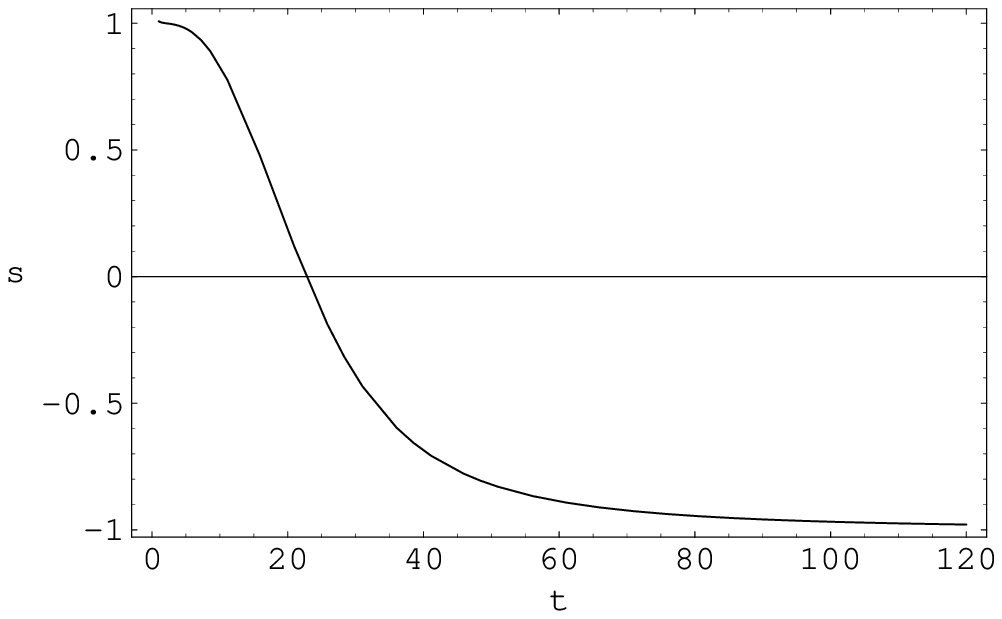}}
\end{center}
\caption{\label{fig1}The modulus high-curvature attractor, for $\beta=10^{-3}$, $\phi_0=6.8$, and $\dot\phi_0\approx 0.7$. Top: $\epsilon^{-1}\sim\s_1$; centre: $\dot\phi/(\epsilon H)\sim \s_2$; bottom: the ghost constraint SC2.}
\end{figure}

Figure~\ref{fig1} shows the behaviour of $\epsilon^{-1}$ and $\dot\phi (\epsilon H)^{-1}$ and the convergence to the attractor at $\epsilon^{-1}\sim Ht \sim \s_1$ and $\dot\phi (\epsilon H)^{-1}\sim\dot\phi\, t\sim\s_2$. The condition SC2 is not satisfied at late times.
The numerical analysis unravels a hidden issue for the asymptotic solutions. The ghost constraints Eqs.~\Eq{gcasy} are valid only at late time and do not say anything about the early-time behaviour of solutions approaching the power-law attractor. Then one might encounter unstable modes even before the attractor is reached. This means that, of all the possible real initial conditions, some are excluded by the presence of ghosts, even those that would have led to an attractor in the phase space.

Finally, an exact unstable solution is $\s_1\approx 0.21$, $\s_2=2$, $\xi_0=54.12$ \cite{ART}. The TT conditions hold but SC2 does not, the latter being $s_\textsc{sc}\approx-2.29<0$, and the solution is unstable also against inhomogeneous perturbations.

%%%%%%%%%%%%%%%%%%%%%%%%%%%%%%%%%%%%%%%%%%%%%%%%%%%%%%%%%%%%%%%%%%%%%%%%%%%%%%%%%%%%%%%%%%%%%%%%%%%%%%%%%%%%%%%%%%%%%%%%

\subsection{The Nojiri--Odintsov--Sasaki modulus}

This model for dark energy (\cite{NOS}, hereafter NOS) is defined by making the choice
\ba
f_1&=&1,\qquad \omega=\pm 1\ ,\\
f_2&=&f_0\,\rme^{2\phi/(\alpha\phi_1)}\\
V&=&V_0\,\rme^{-2\phi/\phi_1},
\ea
where $f_0,\alpha,\phi_1,V_0$ are all constants. By choosing an appropriate value for $V_0$ it is possible to have an accelerating universe today. Let us study the presence of ghosts in such a model. It is useful, before performing a full numerical simulation, to have a look at the initial conditions which determine the parameter space for which there are ghosts. In other words, the conditions TT1, TT2, SC1, and SC2 can be studied for today's initial conditions, which are functions of just $\phi_0$.

\begin{figure}[ht]
\centering
{\psfrag{phisphiu}{$\phi_0$}
\psfrag{SC2m}[][][1][-90]{$s_\textsc{sc}^-\;\;$}
\includegraphics[width=6.5cm]{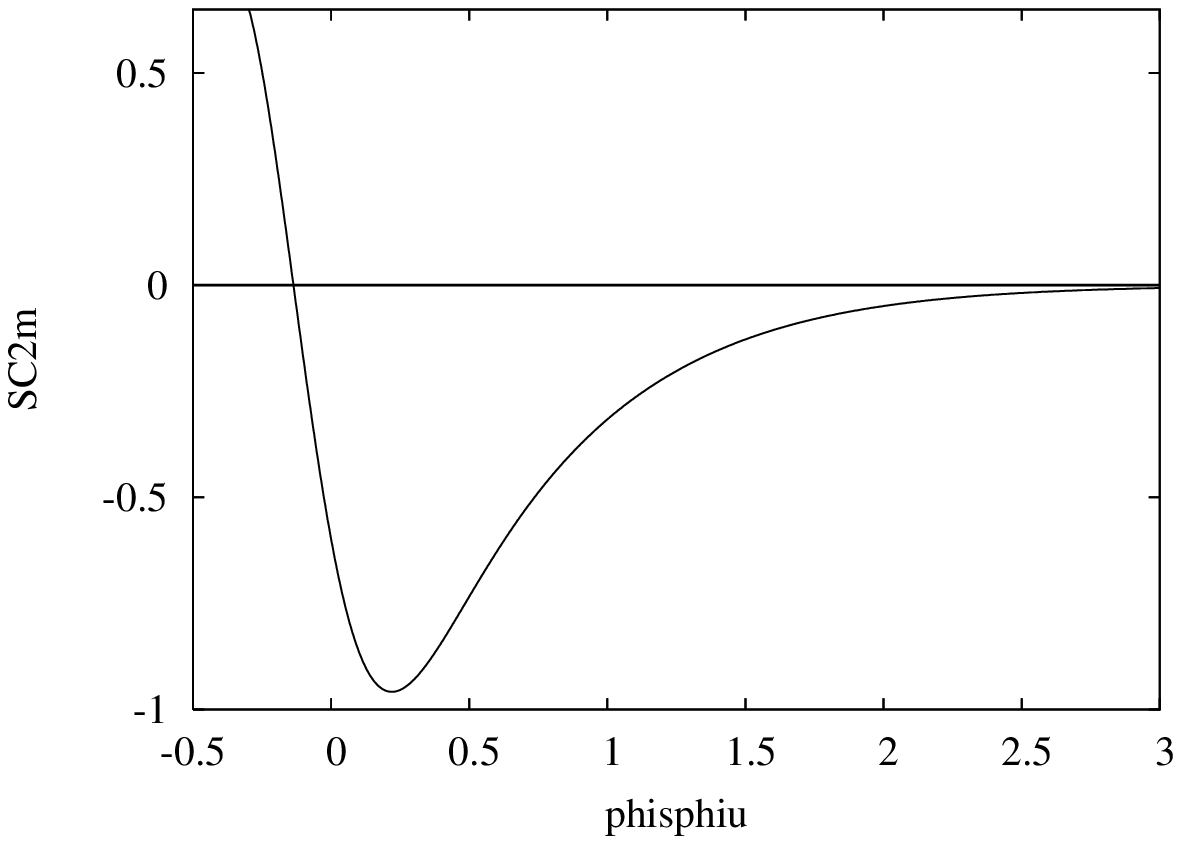}}
{\psfrag{phisphiu}{$\phi_0$}
\psfrag{SC2m}[][][1][-90]{$s_\textsc{sc}^+\;\;$}
\includegraphics[width=6.5cm]{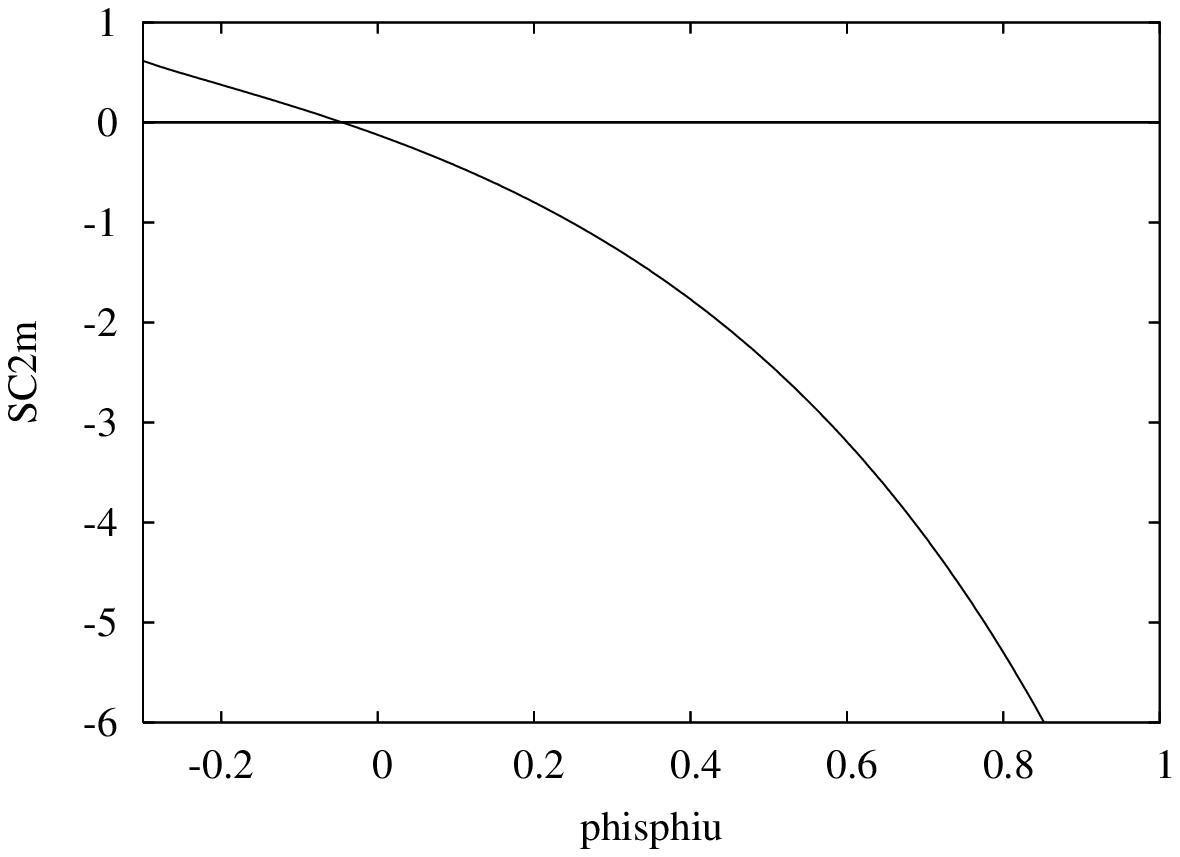}}
\caption{\label{fig2}No-ghost condition SC2 for initial conditions $\phi_0$ in the NOS model. Left: $\dot\phi_0^-$ branch; right: $\dot\phi_0^+$ branch.}
\end{figure}

\begin{figure}[ht]
\centering
{\psfrag{tau}{$\tau$}
\psfrag{ieps}[][][.8][-90]{$\dfrac1\epsilon\;$}
\includegraphics[width=8truecm]{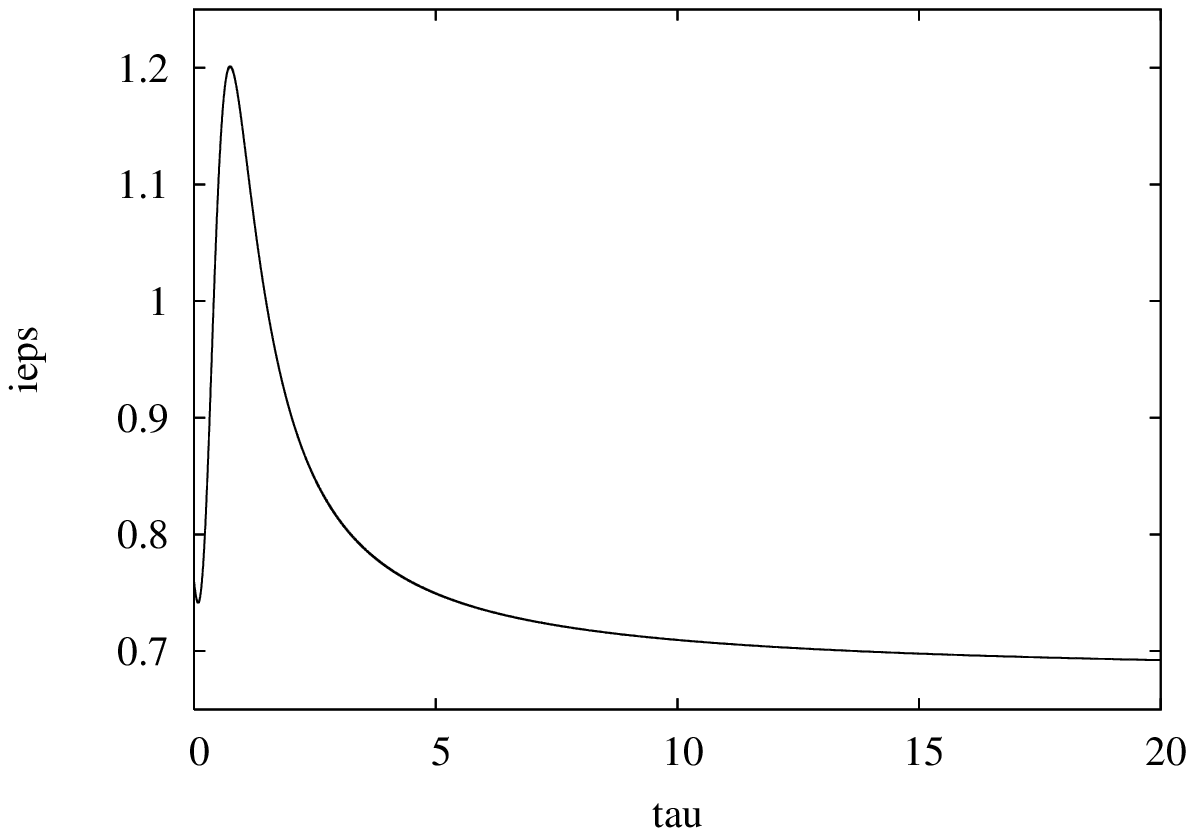}}
{\psfrag{tau}{$\tau$}
\psfrag{tphiHe}[][][.8][-90]{$\dfrac{\dot\phi}{H\,\epsilon}\;\;$}
\includegraphics[width=8truecm]{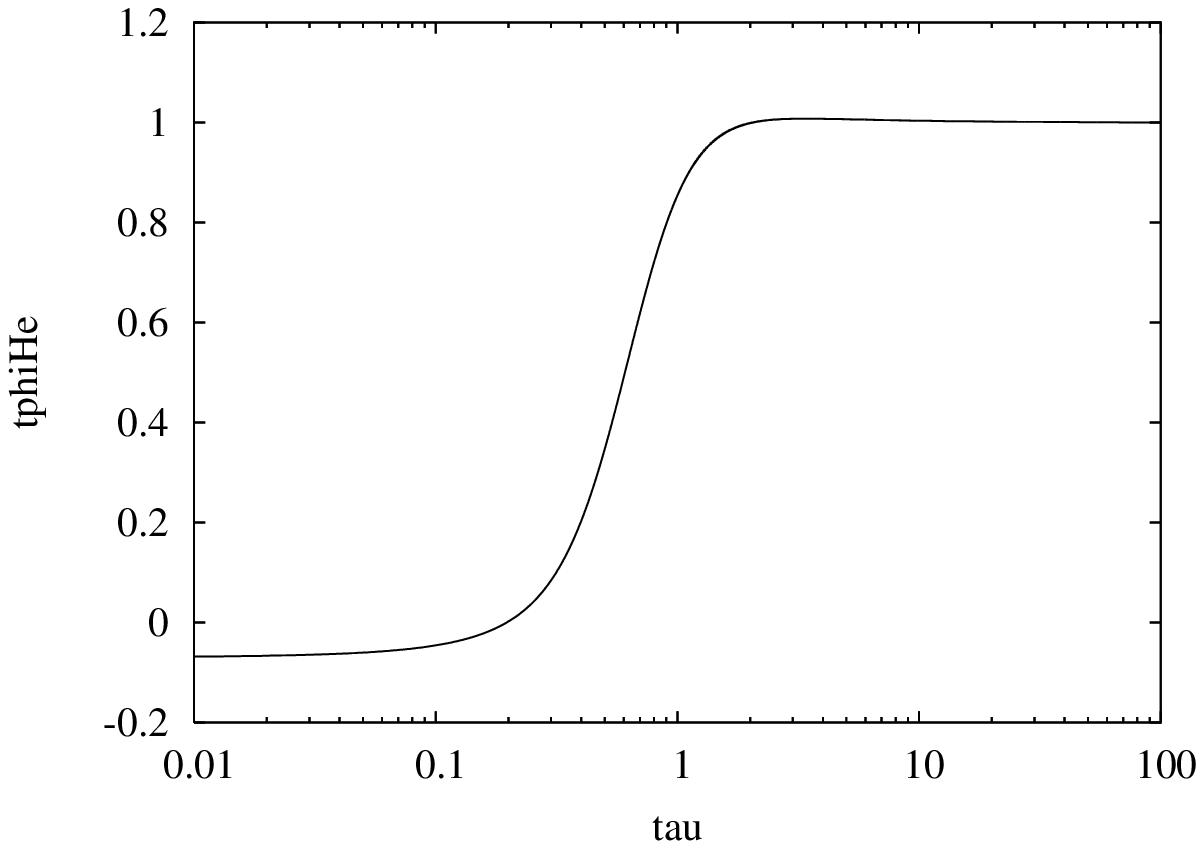}}
{\psfrag{tau}{$\tau$}
\psfrag{SC2}[][][1][-90]{$s_\textsc{tt}$, $s_\textsc{sc}\;\;\;\;\;$}
\includegraphics[width=8truecm]{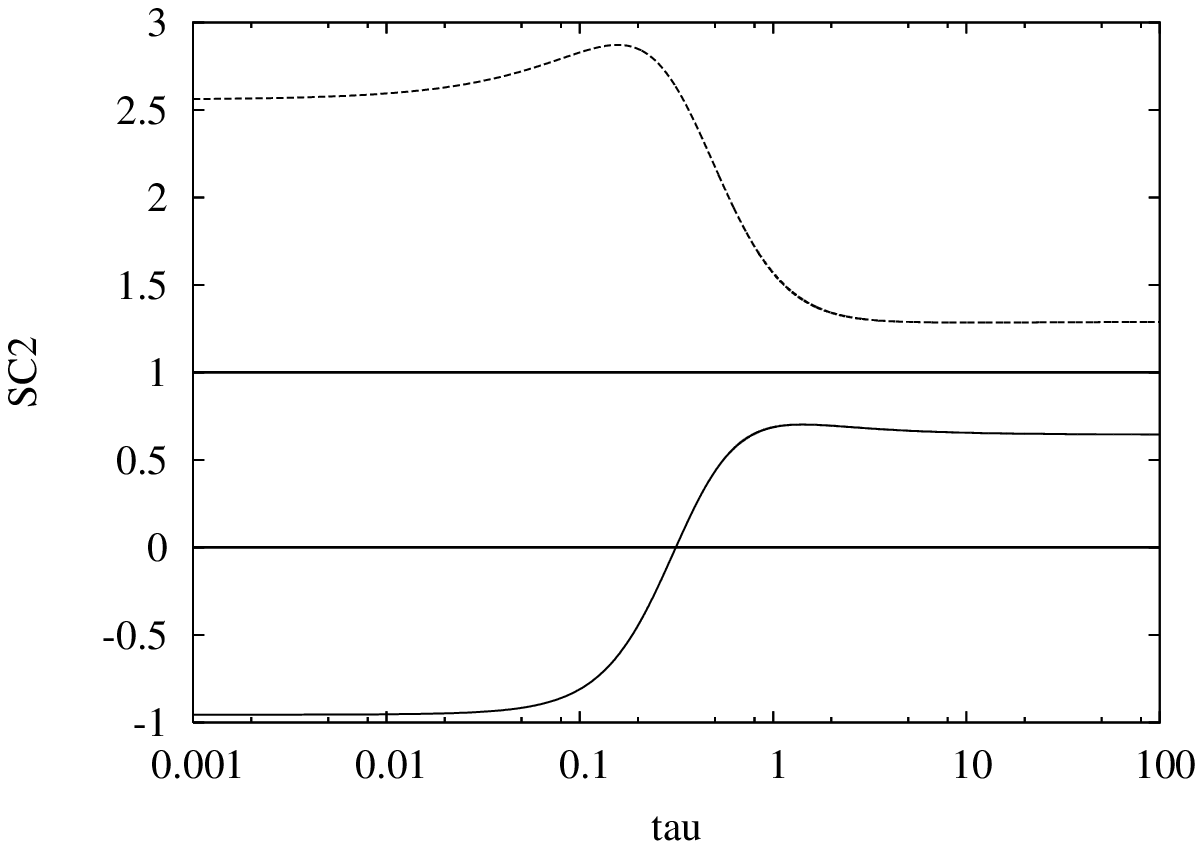}}
\caption{\label{fig3}Late-time behaviour for $\epsilon^{-1}\sim\s_1$ (top), $\dot\phi/(\epsilon H)\sim \s_2$ (central), and the speed of propagation for the scalar (solid) and tensor (dashed) modes (bottom) in the NOS model.}
\end{figure}

For a choice of the parameters of order unity ($\beta=\alpha=\phi_1=f_0=1$, $\omega=1$, and $V_0=0.7$), the reality of $\dot\phi_0$ requires that $\phi_0\gtrsim -0.3$ for both the $\dot\phi_0$-branches. In this case it is possible to check the no-ghost conditions (again, SC1 is trivial) and we find that they are all satisfied but SC2 (negative squared propagation speed, see Fig.~\ref{fig2}) and SL1.
This is consistent with the numerical analysis made by choosing the negative branch and $\phi_0=0.2$. In this case there is a late-time power-law attractor with $\phi\sim\ln t$, and both SC2 and SL1 are not satisfied (see Fig.~\ref{fig3}).

%%%%%%%%%%%%%%%%%%%%%%%%%%%%%%%%%%%%%%%%%%%%%%%%%%%%%%%%%%%%%%%%%%%%%%%%%%%%%%%%%%%%%%%%%%%%%%%%%%%%%%%%%%%%%%%%%%%%%%%%

\subsection{Asymptotic dilaton scenarios}

The logarithmic dilaton with no potential \cite{GV} has a stable, low-curvature asymptotic solution $\s_1=1/\sqrt{3}\approx 0.58$, $\s_2=\sqrt{3}-1\approx 0.73$ (there is no expanding solution in the presence of a cosmological constant). Terms in $f_2$ can be neglected and it is easy to see that the no-ghost and subluminal conditions are trivially satisfied. 

%%%%%%%%%%%%%%%%%%%%%%%%%%%%%%%%%%%%%%%%%%%%%%%%%%%%%%%%%%%%%%%%%%%%%%%%%%%%%%%%%%%%%%%%%%%%%%%%%%%%%%%%%%%%%%%%%%%%%%%%

\subsection{Linear dilaton}\label{lid}

An exact de Sitter solution of the equations of motion can be obtained when $\dot\phi=vt$ and $V=0$, $a_4=-1$, with $v\approx 1.40$ and $H\approx 0.62$ in the bosonic case (there is no such real solution for $a_4=0$). For the GB parametrization, this solution is stable. The no-ghost tensor conditions are satisfied, TT1 being trivial and $Q_2\propto 1-Hv\approx 0.13>0$. SC1 is satisfied, too, but SC2 is violated, being $s_\textsc{sc}\approx -7.61$. In \cite{CHC} the instability arising when $s_\textsc{sc}<0$ was already noted, and interpreted as a breakdown of the linear theory. The same authors claimed also (only as a preliminary result) that the inclusion of higher-derivative terms in the action would flip the sign of $s_\textsc{sc}$. However, the linear expansion of the metric is typically taken as sufficient to determine the viability of a noneffective quantum field theory from the point of view of unitarity, which is also our perspective.

%%%%%%%%%%%%%%%%%%%%%%%%%%%%%%%%%%%%%%%%%%%%%%%%%%%%%%%%%%%%%%%%%%%%%%%%%%%%%%%%%%%%%%%%%%%%%%%%%%%%%%%%%%%%%%%%%%%%%%%%

\subsection{Cyclic scenarios}

Another case in which the action \Eq{act} plays an important cosmological role is in cyclic or bouncing scenarios, in which a contracting phase before inflation can leave an imprint into the cosmic microwave background (CMB) perturbations (e.g., \cite{TBF} and references therein).

\begin{figure}
{\psfrag{t}{$\tau$}
\psfrag{a}{$a$}
\includegraphics[width=7cm]{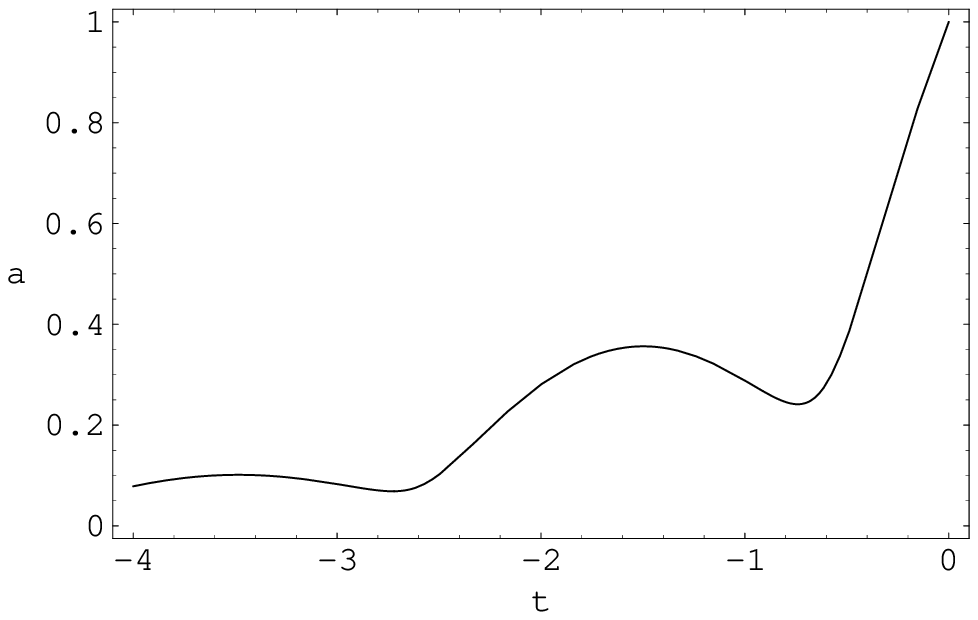}}
{\psfrag{t}{$\tau$}
\psfrag{q}{$q$}
\includegraphics[width=7truecm]{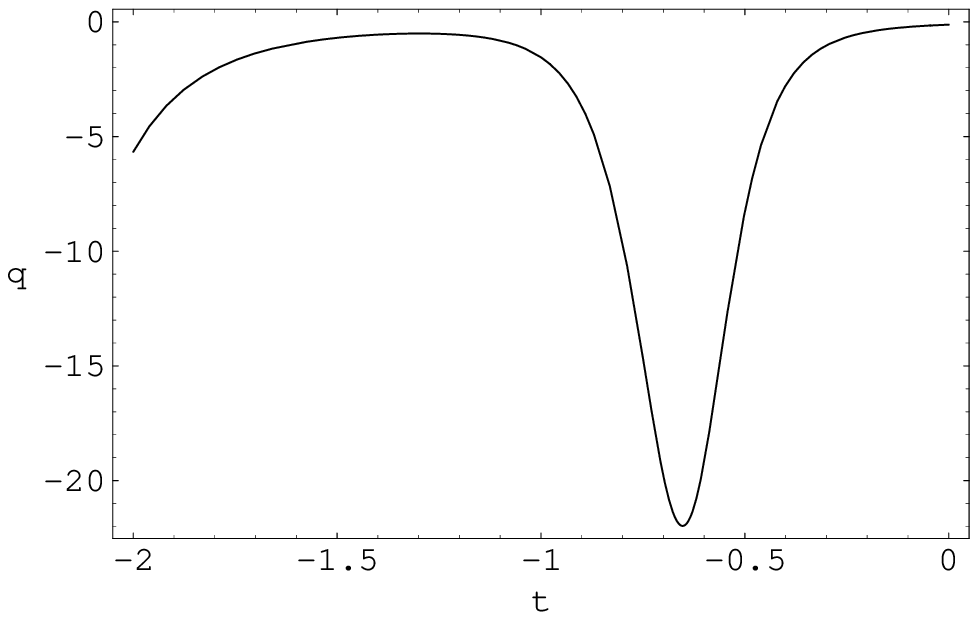}}
\caption{\label{fig4}The scale factor and the function $q$ of the dilaton cyclic solution described in the text, for $\beta=10^{-3}$, $\vp_0=1.8$, and $\dot\vp_0\approx 1.2$. The ghost constraint SC1, which is shown in one cycle, is never satisfied.}
\end{figure}

As an example, we take a dilatonic scalar field with $a_4=0$ and a potential corresponding, in the Einstein frame, to a negative cosmological constant, $\l=1/8$, $\gamma_{41}=\gamma_{i2}=-1$. The scale factor and the condition SC1 are shown in Fig.~\ref{fig4}. We have checked that $|(\Sigma-T)/(\Sigma+T)|<2\times10^{-7}$ and $|y/(\Sigma_{00}+T_{00})|<10^{-11}$. Each bounce is nonsingular in $H$, and there is a correspondent cyclic increase of $|q|$.

We notice that linear theory may break near each bounce \cite{ly02}, and the ghost constraints above might not be too well motivated in this case. However, $a$ is never singular and $q<0$ always, even far from the bounce. The presence of a spin-0 ghost in the general class of bouncing models was already noticed in \cite{AW}, where such instability is regarded as the key ingredient for the bouncing mechanism (however, that setup is rather different from ours). Also, this particular example is expected to be unstable because of both the negative sign of the potential and the choice for $\gamma_{i2}$ (although later on we shall see that instabilities appear anyway for most choices of these coefficients). Here we show it just as another situation in which a non-minimally coupled scalar with a tailored potential and arbitrary initial conditions is plagued by instabilities.
We checked that the background solution does not change qualitatively when $a_4=-1$ (the root closest to $\dot\vp_0$ was taken and found to be almost the same). The ghost constraint SC1 is worsened by the extra term in $a_4$, which is negative definite.

%%%%%%%%%%%%%%%%%%%%%%%%%%%%%%%%%%%%%%%%%%%%%%%%%%%%%%%%%%%%%%%%%%%%%%%%%%%%%%%%%%%%%%%%%%%%%%%%%%%%%%%%%%%%%%%%%%%%%%%%

\subsection{Homogeneous vs cosmological perturbations}

To summarise, we have given examples of stable phase-space solutions which have ghosts or other tree-level instabilities. Looking at the linearly perturbed equations of motion in phase space \cite{CTS}, one can see that the characteristic equation for the mass matrix has a structure which cannot reproduce that of the no-ghost constraints. While a phase-space analysis can track solutions which are stable in the classical sense, the structure of cosmological perturbations is in closer relation with stability at the quantum level, and the two are independent in all respects.

So far we have limited the discussion to an action with the GB parametrization, but solutions (attractor or not) are known also in the general case where the Riemann invariants $R^2$, $P$, and $Q$ have arbitrary coefficients \cite{CTS}. The perturbed equations of motion would be very complicated but it is clear that there would be always a spin-2 ghost in the spectrum of the theory (this is trivial in a Minkowski background).  In fact, for the tensor modes in FRW, the Lagrangian
\be
f_1(\phi)R+f_2(\phi)(a_1R^2+a_2P+a_3Q),
\label{quaL}
\ee
would give rise to a combination $f_1\Box h_i{}^j+u(a_i,f_i,\dot f_i,H,\dots)\Box^2h_i{}^j$ for some function $u$, which has always a ghost unless either $f_1$ or $u$ vanishes identically at all times.\footnote{Higher-derivative terms are usually responsible for violation of unitarity. Taking the example of a scalar field with equation $(\Box+\gamma\Box^2)\phi=0$ and calculating the propagator $G(-k^2)$ in momentum space, one sees that $-G(-k^2)=k^{-2}-(k^2-\gamma^{-1})^{-1}$. Then the particle content of this model is a massless scalar and a ghost scalar with mass $\gamma^{-1}$, the relative $-$ sign between the propagators being the origin of negative-norm (or negative-energy) states. See, e.g., \cite{smi04} for a short introduction to ghosts in higher-derivative theories.} While $f_1\neq 0$ is phenomenologically necessary, the function $u$ cannot vanish for arbitrary choices of $a_i$ and $f_i$ (when $\dot{f}_i=0$, $u\propto a_2+4a_3$ and the coefficients $a_i$ are fixed to the GB choice \cite{zwi}). Then all cosmological solutions of the above generalized quadratic Lagrangian, Eq.~\Eq{quaL}, have ghosts, even if there are attractors.

One can conclude that \emph{the studies of the homogeneous and cosmological perturbations are inequivalent}.

%%%%%%%%%%%%%%%%%%%%%%%%%%%%%%%%%%%%%%%%%%%%%%%%%%%%%%%%%%%%%%%%%%%%%%%%%%%%%%%%%%%%%%%%%%%%%%%%%%%%%%%%%%%%%%%%%%%%%%%%%%%%%%%%%%%%%%%%%%%%%%%%%%%%%%%%%%%%%%%%%%%%%%%%%%%%%%%%%%%%%%%%%%%%%%%%%%%%%%%%%%%%%%%%%%%%%%%%%%%%%%%%%%%%%%%%%%%%%%%%

\section{Towards observationally and theoretically viable solutions}\label{pscon}

So far we have shown that the presence of instabilities depends on the boundary conditions (hereafter `b.c.') selected for each model. It is now time to ask whether there exists a nontrivial region in the parameter space of the b.c. such that the resulting evolution is free from instabilities and compatible with supernov\ae\ observations. The task is hard since the no-ghost and sub-luminal constraints are time dependent and they should be evolved for every b.c.\ and checked to hold at any time throughout the whole evolution. Here we do not fully address this problem but note that a necessary although not sufficient condition to have experimentally viable and (quantum mechanically) stable solutions is to impose the no-ghost and sub-luminal constraints, together with the present bound for $\ddot a$, to be satisfied today. Of course this does not guarantee ghost freedom always, but we will see that the parameter space shrinks considerably nonetheless.

In our particular class of models the parameter space $I$ consists of 17 elements, $I=\{\beta,\,\gamma_{i\!j},\,a_4,\,\vp_0,\,\dot\vp_0\,\,\ddot\vp_0,\, a_0,\,\dot a_0,\,\ddot a_0\}$. Of these, two are fixed by default ($a_0=1=\dot a_0$ in our units), while $\ddot a_0$, $\ddot\vp_0$, and $\dot\vp_0$ are given by Eqs.~\Eq{acceq}, \Eq{vppeq} (evaluated at $\tau=0$) and \Eq{vpeq}, respectively. We can then specialize to the modulus case with arbitrary potential, leaving the effective GB coupling unspecified. Hence
$$
\{\g_{11},\,\g_{12},\,\g_{21},\,\g_{22},\,\g_{23},\,\g_{31},\,\g_{32},\,a_4\}=\{1,\,0,\,\pm 1,\,\sqrt{3},\,\pm 1,\,3/2,\,0,\,0\}\,. 
$$
There is still freedom in the choice of one of the two branches for $\dot\vp_0^\pm$. We shall take the positive branch $\dot\vp_0^+$ when $f_2>0$ (and vice versa), the negative one being unviable because it is then much harder to satisfy the TT constraints on sufficiently long $\vp_0$-intervals.

The parameter space is then reduced, for each branch, to a two-dimensional submanifold, $I=\{\beta,\vp_0\}$. The other parameters are either fixed by hand or extracted from the equations of motion.

In order to show the impact of each set of equations, we shall adopt a gradual approach by first imposing only the four no-ghost constraints (TT1, TT2, SC1, SC2), then the no-ghost and subluminal constraints (SL1, SL2), and finally the no-ghost and sub-luminal constraints together with the observational bounds for $\ddot a_0$ \cite{WM}
\ba
2\sigma:&\qquad& 0.22\lesssim \ddot a_0\lesssim 1.11\,,\label{app02s}\\
1\sigma:&\qquad& 0.55\lesssim \ddot a_0\lesssim 0.97\,,\label{app01s}
\ea
for a total of maximum 7 constraints.\footnote{The bounds Eqs.~\Eq{app02s} and \Eq{app01s} come from the joint analysis of SNLS \cite{ast05}, SDSS \cite{eis05}, WMAP3 \cite{spe06}, and 2dF \cite{ver02,haw02} data, assuming a nonconstant effective barotropic index for the dark energy equation of state.}
In principle, one may take into account even constraints on $\dddot{a}$, as in Table 5 and Fig.~5 of \cite{WM}. Here we will not need to consider this extra bound anyway.

%%%%%%%%%%%%%%%%%%%%%%%%%%%%%%%%%%%%%%%%%%%%%%%%%%%%%%%%%%%%%%%%%%%%%%%%%%%%%%%%%%%%%%%%%%%%%%%%%%%%%%%%%%%%%%%%%%%%%%%%

\subsection{Modulus case}

\begin{figure}
\bc
{\psfrag{u0}{$\vp_0$}
\includegraphics[width=10cm]{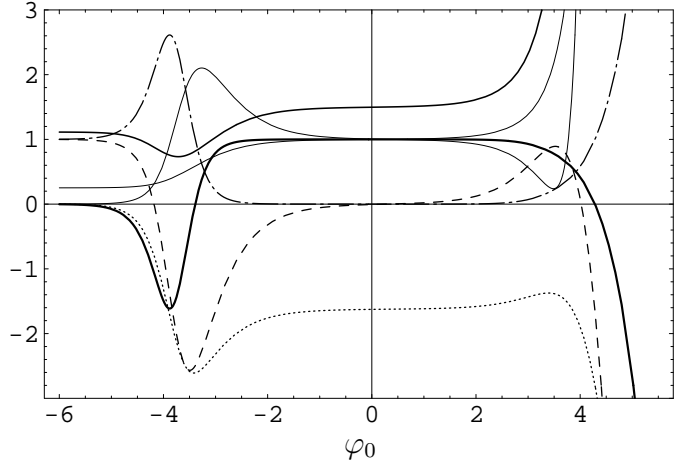}}
\ec
\caption{\label{fig5}No-ghost, sub-luminal, and acceleration conditions for a modulus field with $U=0$ and $\beta=10^{-3}$ as functions of the initial condition $\vp_0^+$ (positive branch). Solid curves with increasing thickness are $Q_1$, $Q_2$, $q$, and $s_\textsc{sc}$. The dashed, dot-dashed, and dotted curves are $1-s_\textsc{tt}$, $1-s_\textsc{sc}$, and $\ddot a_0$, respectively.}
\end{figure}

\begin{figure}
\bc
{\psfrag{u0}{$\vp_0$}
\psfrag{k}[r][][1][0]{$\ln\beta$}
\includegraphics[width=6cm]{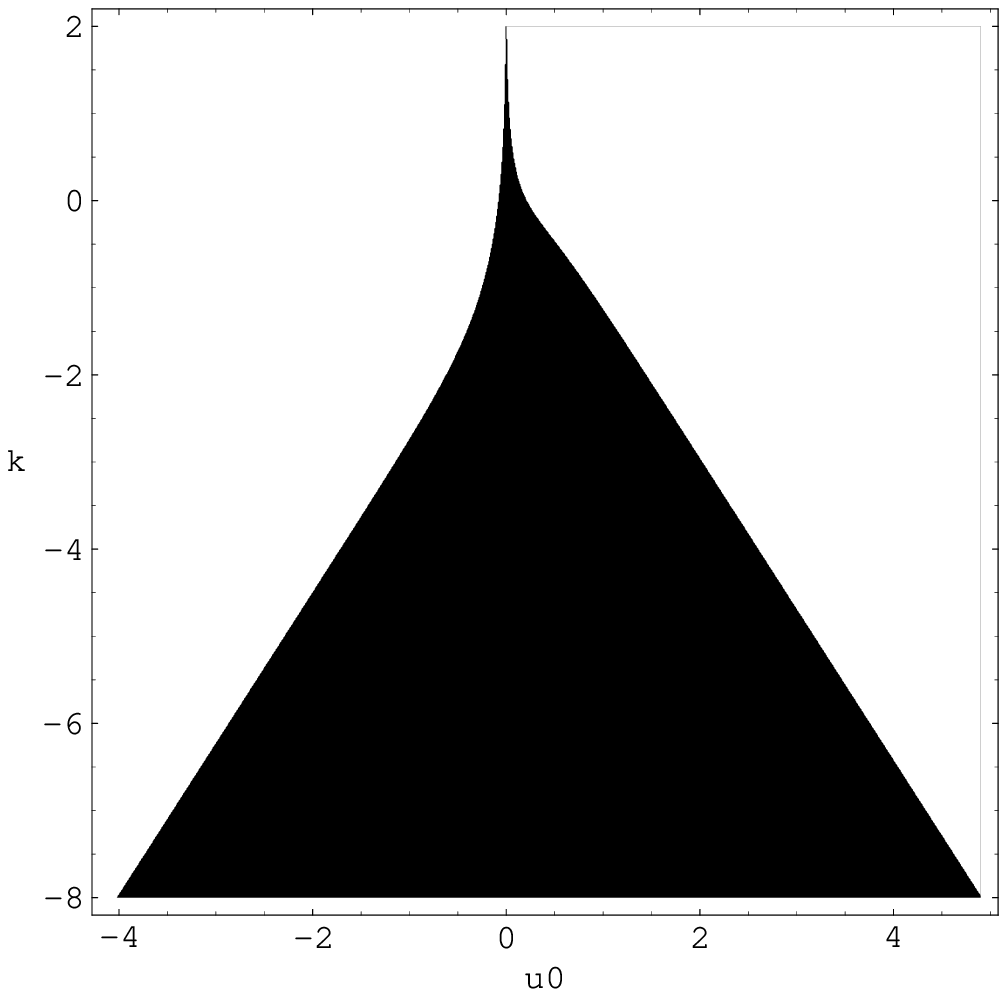}}
{\psfrag{u0}{$\vp_0$}
\psfrag{k}{}
\includegraphics[width=6cm]{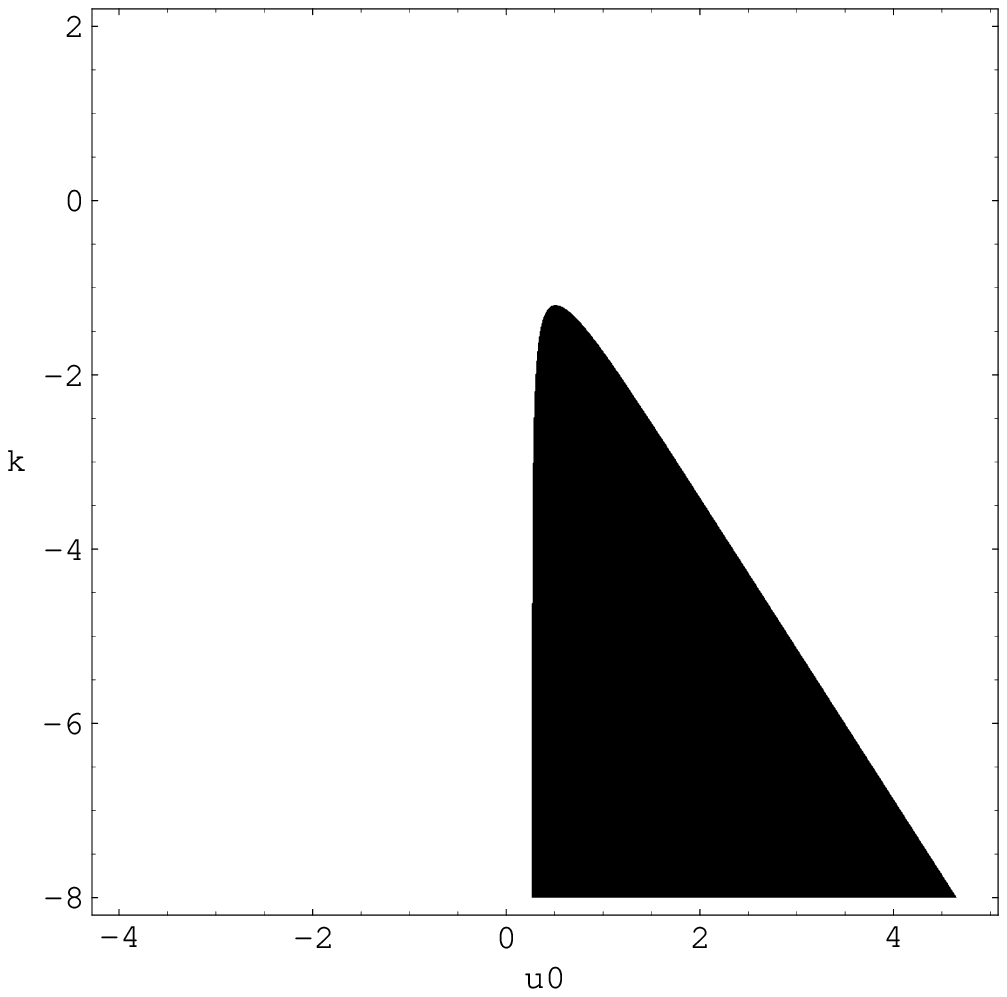}}
\ec
\caption{\label{fig6}Region (black) in the $\vp_0$-$\ln\beta$ plane satisfying the no-ghost (left) and no-ghost + sub-luminal conditions (right), for the positive branch $\vp_0^+$. Here, $U=0$. The model of the previous figure is at $\ln\beta\approx -6.9$.}
\end{figure}

\begin{figure}
\bc
{\psfrag{u0}{$\vp_0$}
\includegraphics[width=10cm]{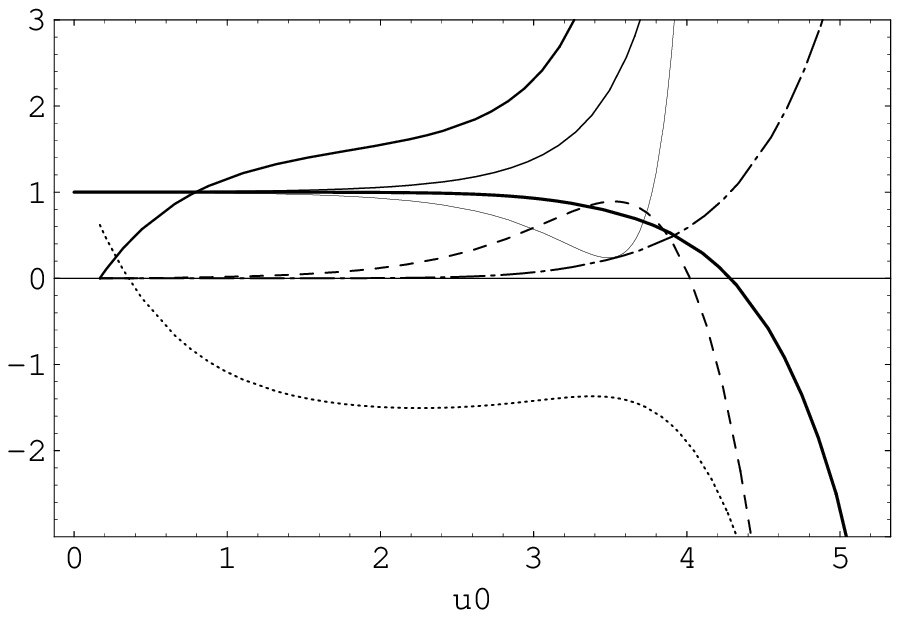}}
\ec
\caption{\label{fig7}No-ghost, sub-luminal, and acceleration conditions for a modulus field with $U=\rme^{-\sqrt{3}\vp}$ and $\beta=10^{-3}$ as functions of the initial condition $\vp_0^+$ (positive branch). Solid curves with increasing thickness are $Q_1$, $Q_2$, $q$, and $s_\textsc{sc}$. The dashed, dot-dashed, and dotted curves are $1-s_\textsc{tt}$, $1-s_\textsc{sc}$, and $\ddot a_0$, respectively.}
\end{figure}

The simplest available example is a modulus in strict sense, that is, with flat potential, $U=0$. In Fig.~\ref{fig5}
all the constraints are shown for a particular value of $\beta$ (see the caption for details). One can then project the intervals, along the $\vp_0$ axis, into which some or all the depicted functions are non-negative, and repeat the process by varying $\beta$. The region in the $\vp_0$-$\ln\beta$ plane where only the no-ghost conditions hold is shown in black in Fig.~\ref{fig6} (left panel). By imposing also the SL conditions, the allowed region is appreciably reduced (right panel), with an upper bound at $\beta\approx 0.30$. This reduction comes actually from SL1 only (tensor modes do not propagate faster than light), being $s_\textsc{sc}\leq 1$ at all times. However, the universe does not accelerate today, $\ddot a_0<0$, and the model can be discarded.
%%%%%%%%%%%%%%%%%%%%%%%%
%%%%%%%%%%%%%%%%%%%%%%%%
%%%%%%%%%%%%%%%%%%%%%%%%
To get acceleration, we must introduce a nontrivial potential. We can take an inverse exponential, $U=\rme^{-\sqrt{3}\vp}$, as in the NOS case (see Fig.~\ref{fig7}). The positive branch $\dot\vp_0^+$ is real when $\vp_0> 0.167$ or $\vp_0< -7.976$, but in the latter case it is easy to see, with a similar analysis, that the no-ghost constraints cannot be satisfied at the same time and there are always instabilities.

\begin{figure}
\bc
{\psfrag{u0}{$\vp_0$}
\psfrag{k}[r][][1][0]{$\ln\beta$}
\includegraphics[width=6cm]{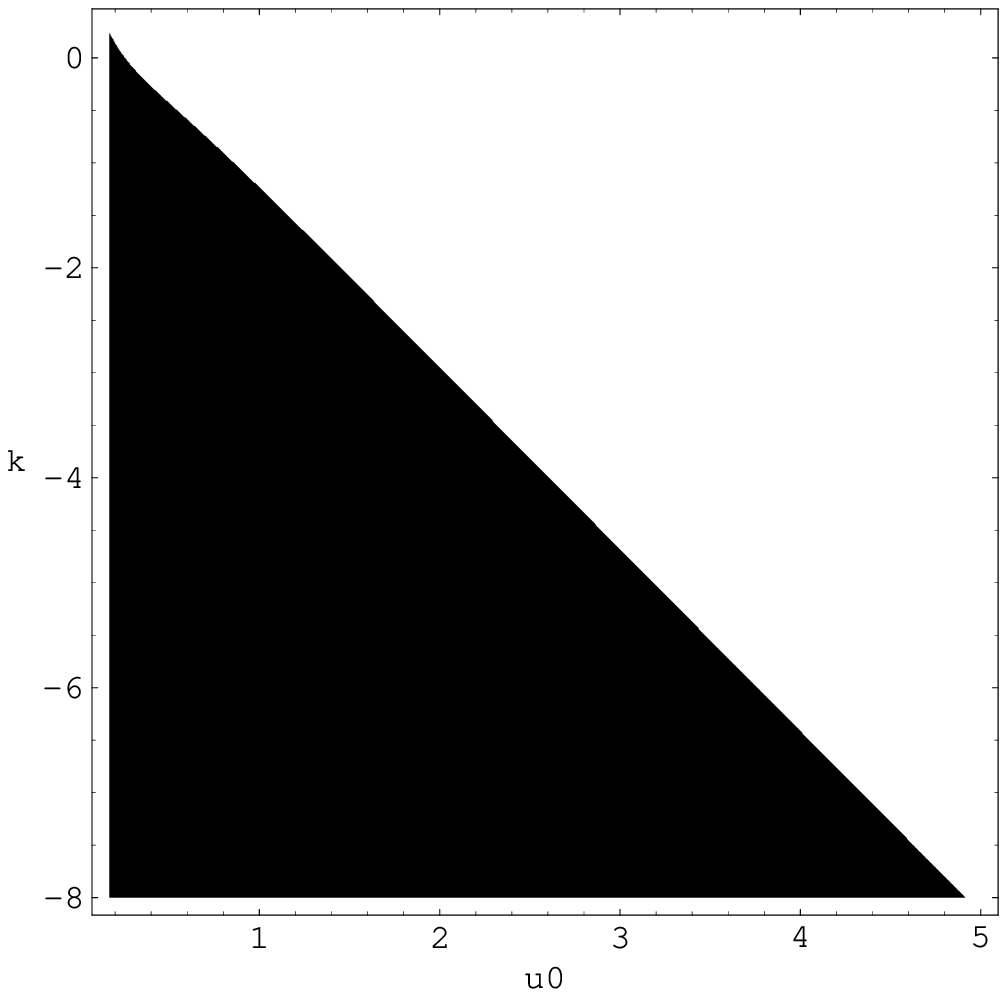}}
{\psfrag{u0}{$\vp_0$}
\psfrag{k}{}
\includegraphics[width=6cm]{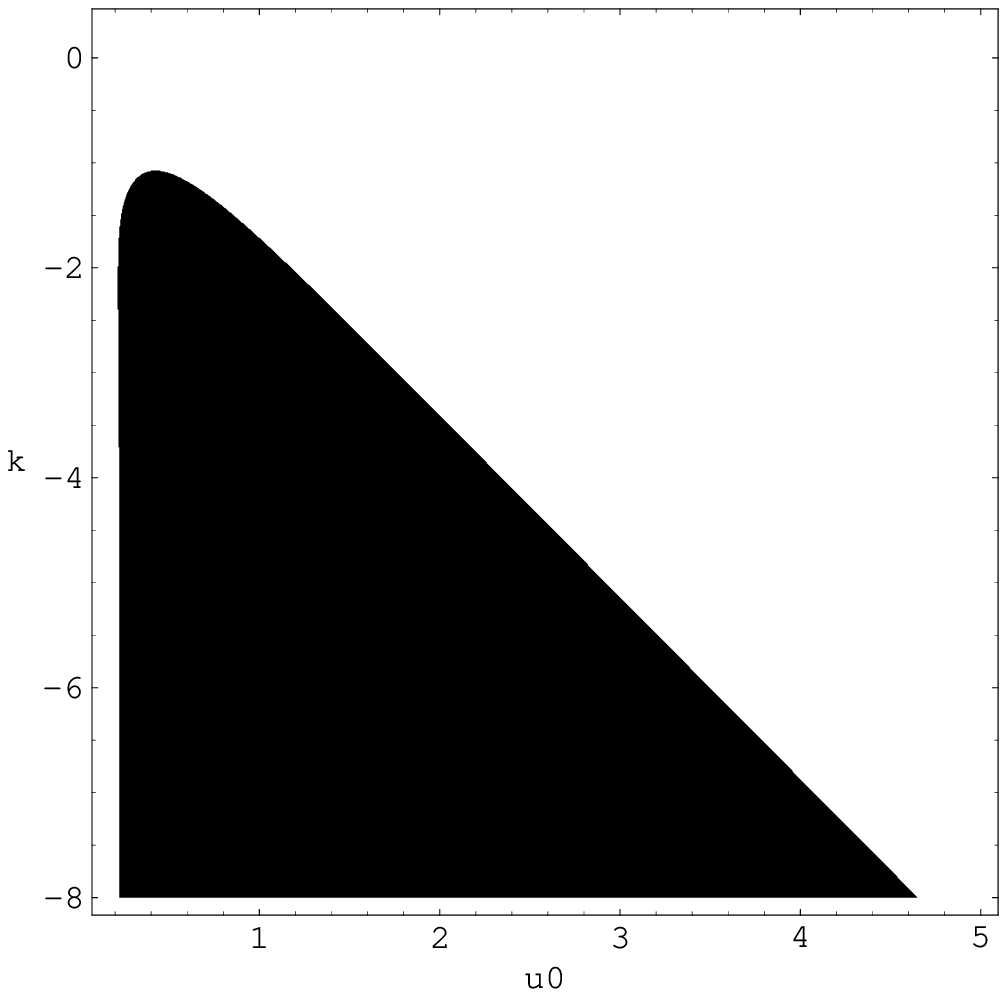}}
\ec
\caption{\label{fig8}Region (black) in the $\vp_0$-$\ln\beta$ plane satisfying the no-ghost (left) and no-ghost $+$ sub-luminal conditions (right), for the positive branch $\vp_0^+$. Here, $U=\rme^{-\sqrt{3}\vp}$. The model of the previous figure is at $\ln\beta\approx -6.9$.}
\end{figure}

\begin{figure}
\bc
{\psfrag{u0}{$\vp_0$}
\psfrag{k}[r][][1][0]{$\ln\beta$}
\includegraphics[width=6cm]{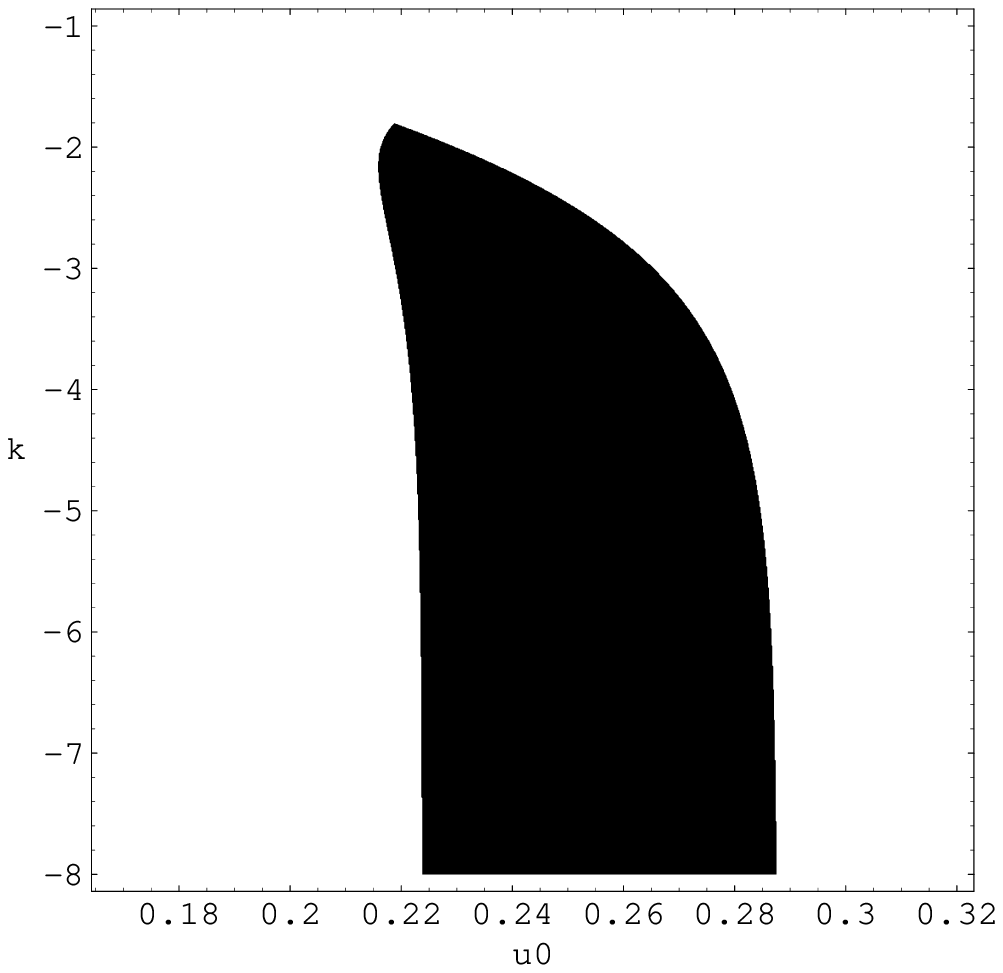}}
\ec
\caption{\label{fig9}Region (black) in the $\vp_0$-$\ln\beta$ plane satisfying the no-ghost, sub-luminal, and acceleration conditions for the positive branch $\vp_0^+$ and modulus potential $U=\rme^{-\sqrt{3}\vp}$. The $2\s$ level of \cite{WM} has been used, Eq.~\Eq{app02s}.}
\end{figure}

Again, SL2 is trivial and the allowed region in the parameter space is mainly reduced by SL1 (Fig.~\ref{fig8}), with $\beta<0.34$. When including Eq.~\Eq{app02s}, only a very narrow strip survives the selection, whose thickness depends on which sigma-level bound one assumes (Fig.~\ref{fig9}). Note that the selection is mainly carried out by the lower bound for $\ddot a_0$. In fact, $\beta> 10^{-2}$ is in the allowed region up to a maximum acceleration $\ddot a_0 \approx 0.41$, so that the model is rejected at the $1\s$ level. Then, when $\ddot a_0$ is increased by $0.01$, there is a leap of 11 orders of magnitude in the GB coupling and the allowed region lies below $\beta\sim 10^{-14}$.
%%%%%%%%%%%%%%%%%%%%%%%%
%%%%%%%%%%%%%%%%%%%%%%%%
%%%%%%%%%%%%%%%%%%%%%%%%
%%%%%%%%%%%%%%%%%%%%%%%%

\begin{figure}
\bc
{\psfrag{u0}{$\vp_0$}
\includegraphics[width=10cm]{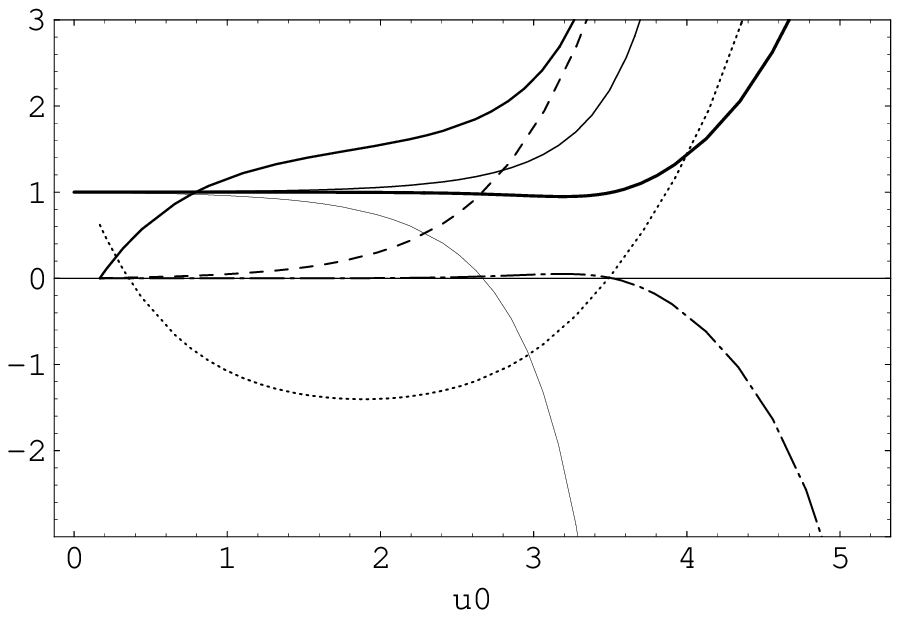}}
\ec
\caption{\label{fig10}No-ghost, sub-luminal, and acceleration conditions for a modulus field with $f_2<0$, $U=\rme^{-\sqrt3\vp}$ and $\beta=10^{-3}$ as functions of the initial condition $\vp_0^-$ (positive branch). Solid curves with increasing thickness are $Q_1$, $Q_2$, $q$, and $s_\textsc{sc}$. The dashed, dot-dashed, and dotted curves are $1-s_\textsc{tt}$, $1-s_\textsc{sc}$, and $\ddot a_0$, respectively.}
\end{figure}

We can learn a few things from these pictures:
\begin{itemize}
\item As one would expect, when $\beta$ decreases and the theory goes towards the GR limit, the allowed intervals on the $\vp_0$ axis for fixed $\beta$ expand.
\item An interesting result is the presence of upper bounds for $\beta$, above which there is at least one moment during the evolution of the universe when instabilities appear. This might mean that the GB Lagrangian, regarded as a perturbative expansion in $\beta$, breaks down and higher-order corrections are necessary for consistency.
\item In at least one example the sub-luminal condition, SL1, removes most of the parameter space. From Fig.~\ref{fig5} it is clear that, keeping only the solid lines, the allowed interval along $\vp_0$ is rather larger than when one imposes also the SL constraints. This is confirmed in the left plot of Fig.~\ref{fig6}, where the positive branch allows also negative values for $\vp_0^+$. Here there is no upper bound for $\beta$, which indeed means that it comes from the requirement of sub-luminal propagation.
\item The condition of $\ddot a_0$ is by far the most stringent bound for the parameter space. In general, not all choices for $\g_{i\!j}$ are compatible with observations.
\end{itemize}
The above features are qualitatively the same when $f_2<0$.\footnote{We have checked that they do not change also for other values of $\gamma_{i2}$, although the allowed parameter space changes and may be constituted, for each $\beta$, by the union of disconnected intervals along $\vp_0$-slices.} In this case, the negative branch $\vp_0^-$ is the viable one; the example $U=\rme^{-\sqrt3\vp}$ with the sign of $f_2$ flipped is shown in Fig.~\ref{fig10}. Again, the allowed region is for $\beta<0.30$.
%%%%%%%%%%%%%%%%%%%%%%%%
%%%%%%%%%%%%%%%%%%%%%%%%

\begin{figure}
{\psfrag{t}{$\tau$}
\psfrag{Ht}[][][.8][0]{$\dfrac1\epsilon$}
\includegraphics[width=6.8cm]{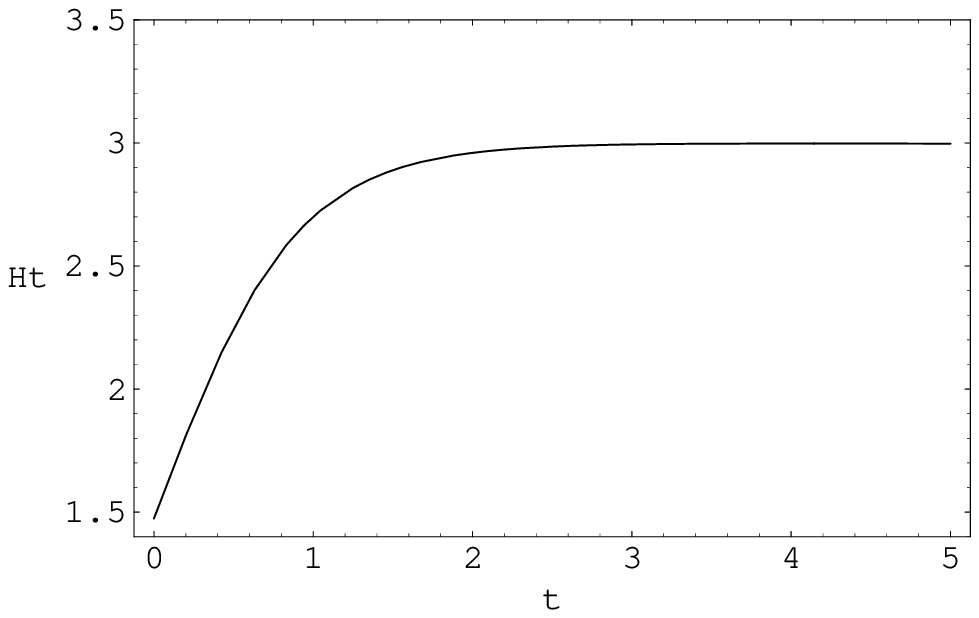}}
{\psfrag{t}{$\tau$}
\psfrag{ut}[][][.8][0]{$\dfrac{\dot\vp}{H\epsilon}$}
\includegraphics[width=6.8cm]{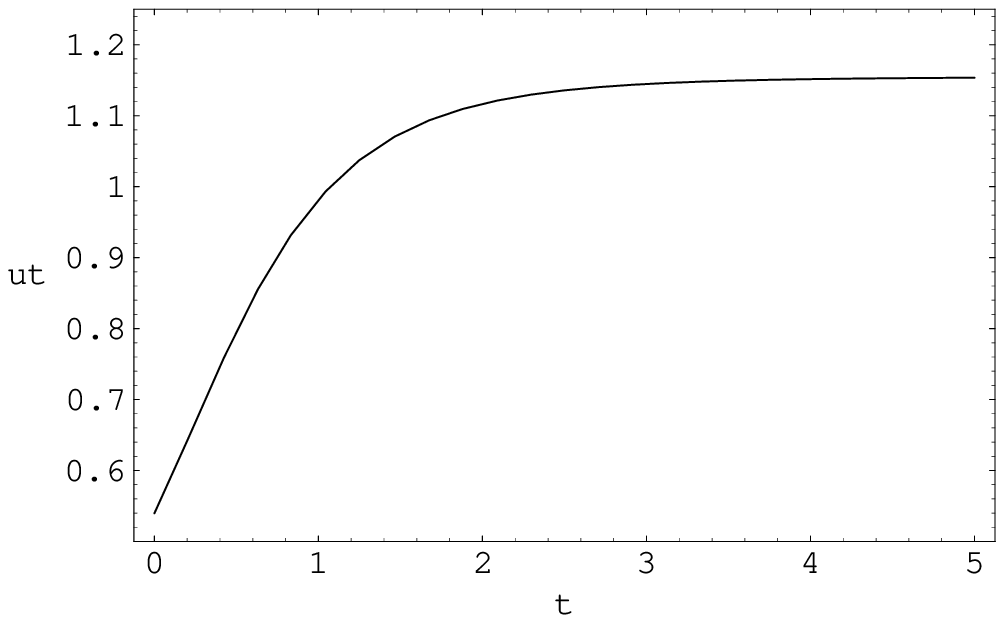}}
\caption{\label{fig11} Future evolution of $\epsilon^{-1}\sim\s_1$ (left) and $\dot\vp/(\epsilon H)\sim \s_2$ (right) for a modulus with $\beta=10^{-3}$, $\vp_0=0.25$, and $U=\rme^{-\sqrt{3}\vp}$.}
\end{figure}

Let us take $\vp_0=0.25$ in the example of Fig.~\ref{fig7} (so that $\ddot a_0\approx 0.32$). At late times one recovers an asymptotic evolution as given by Eq.~\Eq{logas} with $\s_1\sim 3$ and $\s_2\sim 2/\sqrt{3}\approx 1.15$ (see Fig.~\ref{fig11}). With $\s_2=2/\sqrt{3}$ the contribution of the potential goes as the scalar kinetic energy and the Hubble parameter, $\dot\vp^2\sim H^2\sim t^{-2}$.
%%%%%%%%%%%%%%%%%%%%%%%%
%%%%%%%%%%%%%%%%%%%%%%%%
Defining $\Delta^{(0)}\equiv |H-x^{(0)}|/H$ and $\Delta^{(1)}\equiv |H-x|/H$, where $x$ is given by Eq.~\Eq{x}, one has that $\Delta^{(0)}<4\times 10^{-4}$ and $\Delta^{(1)}<4\times 10^{-7}$ around $z\lesssim 1$. The approximated solutions deviate more than $1\%$ from the numerical one when $\Delta^{(0)}> 10^{-2}$ for $z\gtrsim 1.6$ %1.748
and $\Delta^{(1)}>10^{-2}$ for $z\gtrsim 2.5$. %2.485

As anticipated, this model-selection procedure is not sufficient by itself to obtain solutions which are always stable.
It turns out that $s_\textsc{tt}$ becomes larger than 1 when $z\gtrsim 0.2$ and is sub-luminal afterwards. That happens well before (going to the past) the point from which the numerical solution $H$ and the approximated GR one $x^{(0)}$, Eq.~\Eq{xappr0}, begin to diverge. That at early times one does not get an Einsteinian evolution is confirmed by a high-redshift inspection of $\vp$, which is not constant.

The rising of superluminal tensor modes is associated with an increase of the field acceleration, since $s_\textsc{tt}\sim -\ddot\vp/(H\dot\vp)$. We have checked numerically that, when $s_\textsc{tt}\gtrsim 1$, in general there is a corresponding bump in $\dot\vp$. Hence, one way to avoid superluminality  is achieved by the field slowly rolling down its effective potential. In fact, $-\ddot\vp/(H\dot\vp)$ is nothing but the definition of the second slow-roll parameter $\eta_{\rm SR}$ which, in this form, is independent from the type of scalar equation of motion. In the theory under study, $\vp$ is non-minimally coupled and, in general, slow rolling ($\eta_{\rm SR} \lesssim 1$) will not be associated with a flat potential. Another possibility might be to have an effective potential with a false vacuum in which $\vp$ sits for sufficiently long time (i.e.\ with tunneling decay rate $\sim H^{-1}$) to allow for a standard GR evolution. At some point the field tunnels down to the true vacuum through a potential barrier $\Delta U=O(\beta)$, and the GB term becomes dynamical (and, in the best case, also dynamically important).

Both these scenarios require a complicated effective potential and a highly nontrivial dynamical behaviour. Without a more solid theoretical implementation at hand, we can regard them only as proposals for future directions.
One possibility would be to consider flux compactification models~\cite{GKP}, where the dynamics are given in terms of polynomial interactions between the different moduli (plus, possibly, nonperturbative effects). Those would give rise to more involved dynamics than the models presented here, and false vacua would almost certainly exist. It must be noted that, in its vast majority, these models have been constructed in the context of type IIB string theory, for which a GB term is not realised. However, analogous constructions have been proposed in the context of heterotic string~\cite{GKLM,GLM,DGLM}, and it would be interesting to study the implication of the presence of flux in the dynamics of extended gravity models.

%%%%%%%%%%%%%%%%%%%%%%%%%%%%%%%%%%%%%%%%%%%%%%%%%%%%%%%%%%%%%%%%%%%%%%%%%%%%%%%%%%%%%%%%%%%%%%%%%%%%%%%%%%%%%%%%%%%%%%%%

\subsection{The pseudo-modulus of the NOS model}

\begin{figure}
\begin{center}
{\psfrag{u0}{$\vp_0$}
\psfrag{k}[r][][1][0]{$\ln\beta$}
\includegraphics[width=6cm]{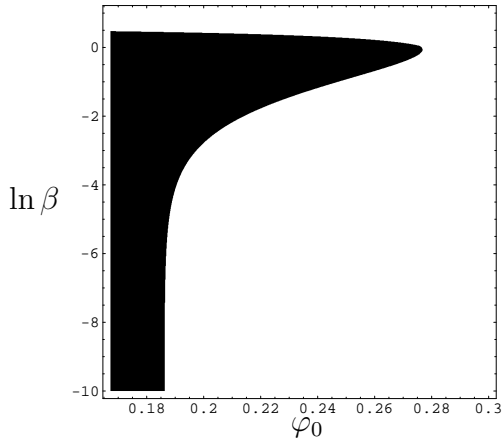}}
\end{center}
\caption{\label{fig12}Region (black) in the $\vp_0$-$\ln\beta$ plane satisfying the no-ghost and sub-luminal conditions for the positive branch $\vp_0^+$ of the NOS model for $f_2=U=\rme^{-\sqrt{3}\vp}$. The contour does not change if acceleration bounds are taken into account.}
\end{figure}

In the model actually considered in~\cite{NOS}, $f_2$ is an exponential instead of a hyperbolic cosine and $\omega=1$. We have checked that even in that case there is a nontrivial parameter space, which is shown in Fig.~\ref{fig12}
%%%%%%%%%%%%%%%%%%%%%%%%
%%%%%%%%%%%%%%%%%%%%%%%%

%%%%%%%%%%%%%%%%%%%%%%%%%%%%%%%%%%%%%%%%%%%%%%%%%%%%%%%%%%%%%%%%%%%%%%%%%%%%%%%%%%%%%%%%%%%%%%%%%%%%%%%%%%%%%%%%%%%%%%%%

\subsection{Dilaton case}

On can perform the same analysis for the dilaton. We have tested a scalar field with $\g_{i2}=-\sqrt{3}$ for (i) $U=0$, (ii) $U=f_1$, and (iii) $U=\omega$. Since none of this cases is robust against theoretical and experimental bounds, as stressed throughout the paper and here reconfirmed, we only quote the main results without producing any figure.

In general, there is a nontrivial parameter space compatible with the four no-ghost constraints. When imposing also the sub-luminal constraints, only case (iii) for the `positive' branch (which does not give a cyclic evolution) survives the selection, due to the fact that $s_\textsc{tt}>1$ typically in these models. When increasing the magnitude of the $|\g_{i2}|$ coefficients there appear regions in which $s_\textsc{tt}<1$, but then $s_\textsc{sc}$ is too large, implying that Eq.~\Eq{brd} is far from guaranteeing sub-luminal propagation. Even examples within the allowed regions, however, are superluminal in the past. Cases (i) and (ii), as well as others we have sampled, never accelerate, as anticipated in Section \ref{numic}.

\subsection{An example of viable model}

One might ask if there is a chance to find a viable model which fits data and satisfies the no-ghost and sub-luminal constraints. Of course, to set $\beta=0$ solves the problem, but this is tantamount to saying that GB actions are not viable in cosmology. However, it would be interesting to find, as in the above analysis, an upper bound for $\beta$ which is physically consistent. We have studied the evolution of these models from a chosen redshift $z_i>0$. In order to do this we have used the number of $e$-foldings $N\equiv\ln(1+z)$ as the new time variable, and $u\equiv\ln (H/H_0)$ and $\varphi$ as dynamical variables. As initial conditions, at $N_i=\ln(1+z_i)$, we have chosen $u_i=\ln(H_\textsc{gr}/H_0)$ and $\varphi$ such that $\ddot a \approx \ddot a_\textsc{gr}$, where the subscript `GR' denotes the general relativistic value of the above quantities. $d\varphi/dN$ is found by using the Friedmann constraint. In this way, the dynamics were initially close to pure GR (universe filled with radiation, dust, and a cosmological constant). For the kinematic evolution of the universe to be acceptable, $\ddot a_0>0$ and $u$ needs to vanish at $N=0$.

\begin{figure}[ht]
\begin{center}
{\psfrag{z}{$z$}
\psfrag{HdivHGR}[][][1][-90]{$\log_{10}\left\vert\tfrac{H-H_\textsc{GR}}{H+H_\textsc{GR}}\right\vert$\qquad\qquad\quad\quad}
\includegraphics[width=8truecm]{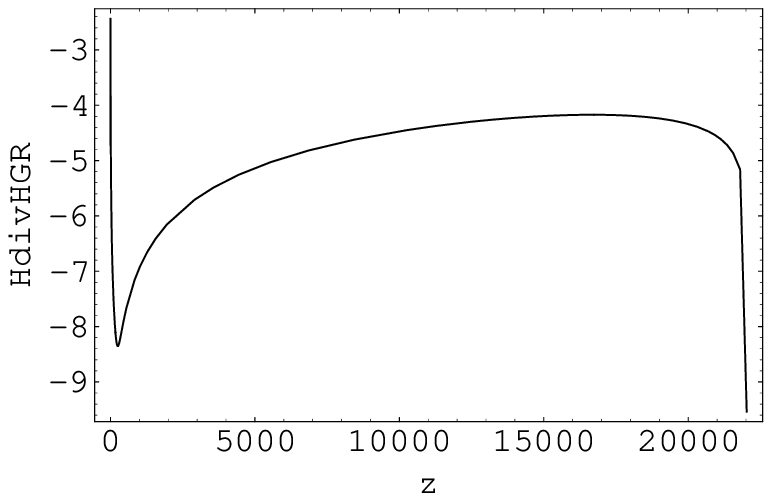}}\\
{\psfrag{z}{$z$}
\psfrag{D2a}[][][1][-90]{$\ddot a$\quad}
\includegraphics[width=8truecm]{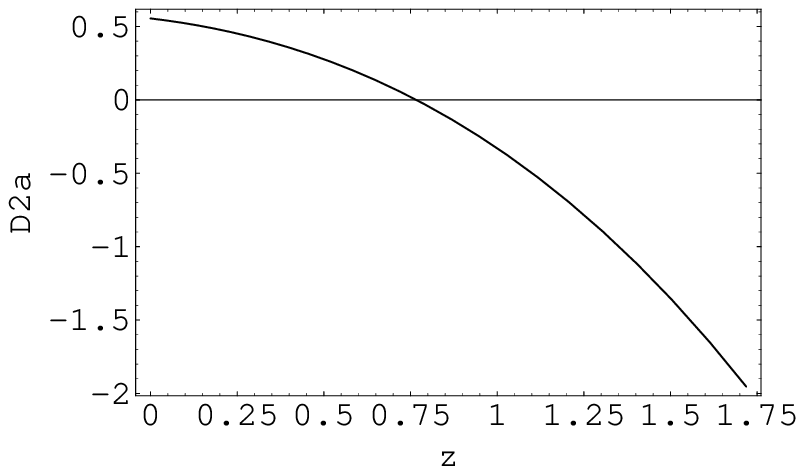}}\\
{\psfrag{z}{$z$}
\psfrag{sTT}[][][1][-90]{$\begin{array}{l}\log_{10}(1-s_\textsc{tt})\\ \log_{10}(1-s_\textsc{sc})\end{array}$\qquad\qquad\qquad}
\includegraphics[width=8truecm]{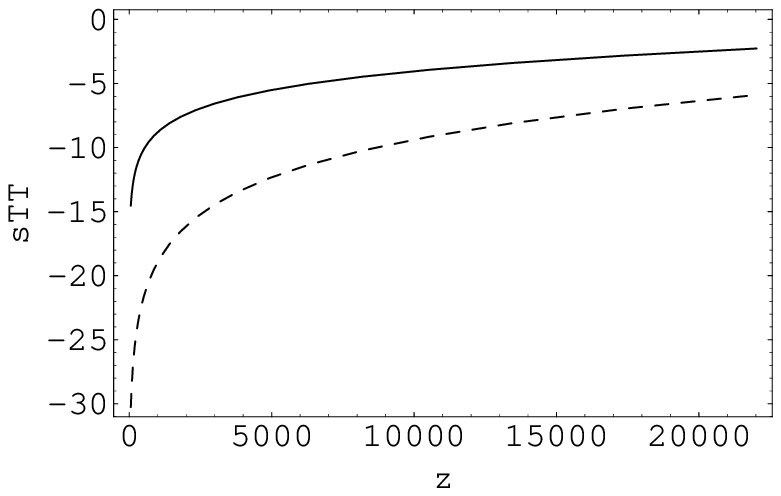}}
\end{center}
\caption{\label{fig13}Comparison of the time-evolution for GR and GB cosmologies (top), $\ddot a$ (central), and the speed of propagation for the scalar (dashed) and tensor (solid) modes (bottom) for the viable example described in the text.}
\end{figure}

For the dilaton ($\gamma_{21}=1/8$, $\gamma_{41}=0.7$, $\gamma_{42}=\pm1$) we found that it is, in general, hard to find good agreement with GR evolution, leading this to the cyclic behaviour already discussed before. On the other hand things change when we consider a modulus-like field. In fact, we have found a model which works, by choosing $f_1=1$, $f_2=\exp(-\sqrt{3}\vp)$, $\gamma_{41}=0.7,\gamma_{42}=1$, $\beta\sim10^{-15}$ and $\varphi_i=0.125$ as initial condition at $N_i=10$ ($z_i\sim2\times10^4$). Then the universe follows a GR evolution, accelerates today and all the no-ghost constraints hold. Both the speeds of propagation differ from one at most by one part in $10^7$ at high redshifts (see Fig.~\ref{fig13} for all these features).
It should be noted that it is crucial to choose the $\dot\varphi^{+}$-branch as initial condition for $\dot\varphi$, otherwise superluminal modes appear at late times. This means that the result is strongly dependent on the initial conditions. Furthermore, as $\beta$ increases, in order to have early-time agreement with GR, also the value for $\vp_i$ needs to increase: for example, for $\beta=10^{-10}$, we require $\vp_i\sim 4.5$ at $z_i$, but today $H$ diverges from $H_0$.

In general, there is an upper value for $\beta_{\rm eff}$, larger than the real coupling $\beta$, below which one recovers, for some models, the Einstein--Hilbert evolution from some redshift on. For other choices of the parameters we have also checked that, when $\beta_{\rm eff}$ is lowered, the onset of superluminality is shifted backward in time and one gets closer and closer to a GR behaviour. However this means that, in order to get a GR evolution at early times (when the contribution of the quadratic curvature term is comparable with the linear EH), the effective coupling should be $\beta_{\rm eff}\approx\beta$. We finally noticed that, if $f_2$ is assumed to be the hyperbolic cosine of the genuine modulus case, superluminal modes appear again for the same initial conditions.

%%%%%%%%%%%%%%%%%%%%%%%%%%%%%%%%%%%%%%%%%%%%%%%%%%%%%%%%%%%%%%%%%%%%%%%%%%%%%%%%%%%%%%%%%%%%%%%%%%%%%%%%%%%%%%%%%%%%%%%%%%%%%%%%%%%%%%%%%%%%%%%%%%%%%%%%%%%%%%%%%%%%%%%%%%%%%%%%%%%%%%%%%%%%%%%%%%%%%%%%%%%%%%%%%%%%%%%%%%%%%%%%%%%%%%%%%%%%%%%%

\section{Including a second scalar field}\label{sec2f}

One may suggest that the approach followed so far may not be sufficient to select models with a runaway scalar coupled to the GB combination because, in general, there will be more than one field having similar couplings. An example can be a complex modulus field instead of a real one. Just to give an idea of what changes when considering two real scalars coupled with a GB term, in this section we shall provide only a partial analysis of such case, concentrating on the behaviour of the tensor modes and giving a numerical example to evaluate the impact of a second field. The answer is that, at least when the extra field has a vanishing potential, its effect is negligible.

We call $\cS_\vp$ the action given by Eq.~\Eq{act} with $f_1=1$, $\omega(\vp)= {\rm const}$, and $a_4=0$. As discussed in \cite{ART,neu06}, one introduces the following extra terms:
\ba
\cS &=&\cS_\vp+\cS_\chi,\\
\cS_\chi&=&\int d^4x\,\sqrt{-g}\left[-\tfrac12\,\vs\,\N_\a\chi\N^\a\chi+f_3(\chi)\,\cL_{\rm GB}\right]\ ,
\label{pane}
\ea
where $\chi$ is another scalar modulus and $\vs$ is constant. According to Eq.~\Eq{pane}, the two scalar fields do not interact at tree level. Then the Einstein equations have now some additional contributions,
\ba
\Xi_{\mu\nu}&=&\Sigma_{\mu\nu}-\vs\N_\mu\chi\N_\nu\chi+g_{\mu\nu}\left(4R\Box f_3-8R^{\a\t}\N_\a\N_\t f_3+\tfrac12\,\vs\N_\a\chi\N^\a\chi\right)\nonumber\\
&&\qquad-4R\N_\mu\N_\nu f_3-8R_{\mu\nu}\Box f_3-8R_{(\mu}{}^{\s\t}{}_{\nu)}\N_\s\N_\t f_3 \nonumber \\
&&\qquad+16R_{\s(\mu}\N^\s\N_{\nu)} f_3\ ,
\ea
and the equation for $\chi$ reads
\be
\vs\Box\chi+f_{3,\chi}\,\cL_{\rm GB}=0\ .
\label{eomchi}
\ee
In a FRW background these equations read, after redefining $\chi\to \sqrt3\,\chi/\kappa$ ,
\ba
&&\beta\,(\dot f_2+\dot f_3)\,H^3+H^2=\rho_m+\rho_r+\tfrac12\,\omega\dot\vp^2+
\tfrac12\,\varsigma\dot\chi^2+U,\label{freqchi}\\
&&\omega(\ddot{\vp}+3H\dot{\varphi})+U_{,\vp}-\beta f_{2,\vp}H^2(H^2+\dot H)= 0,\\
\label{pheom2chi}
&&\vs(\ddot{\chi}+3H\dot{\chi})-\beta f_{3,\chi}H^2(H^2+\dot H)= 0\ .\label{eqchi}
\ea
Since the tensor modes do not mix with scalar ones, the conditions TT1, TT2, and SL1 can then be applied immediately, by replacing $f_2$ with $f_2+f_3$:
\begin{eqnarray}
&& 0<1+\beta\,(\ddot f_2+\ddot f_3)\label{TTnoghostchi}\,,\\
&&0<\frac{1+\beta\,(\ddot f_2+\ddot f_3)}{1+\beta\,H\,(\dot f_2+\dot f_3)}\leq1\ .\label{TTcchi}
\end{eqnarray}
The introduction of a new field enlarges the available parameter space, and a full numerical analysis is beyond the scope of this paper. However, a few insights in this model can be obtained by looking at the particular case
\ba
f_2(\vp) &=&\gamma_{21}\,\exp(\gamma_{22}\,\vp) \,,\\
f_3(\chi) &=&\gamma_{51}\,\cosh(\gamma_{52}\,\chi) \,,
\ea
while $U$ is given by Eq.~\Eq{expot}.
Furthermore, there are three free independent initial conditions, $\vp_0$, $\chi_0$, and $\dot\chi_0$, whereas $\dot\vp_0$ can be found by using the Friedmann constraint~(\ref{freqchi}). By choosing values of the parameters to be of order one ($\gamma_{21}=1$, $\gamma_{22}=2$, $\gamma_{41}=0.7$, $\gamma_{42}=-2$, $\gamma_{51}=\gamma_{52}=1$) and initial conditions $\varphi_0=\chi_0=0$, the field $\chi$ evolves to a constant value, i.e.~$\dot\chi\to0$. This solution is an attractor, because it is reached independently from the initial condition given to $\dot\chi_0$. Therefore, at late times the evolution tends to be the same solution of the NOS case discussed before. In particular, as it happened in the NOS case, the tensor modes are superluminal.

%%%%%%%%%%%%%%%%%%%%%%%%%%%%%%%%%%%%%%%%%%%%%%%%%%%%%%%%%%%%%%%%%%%%%%%%%%%%%%%%%%%%%%%%%%%%%%%%%%%%%%%%%%%%%%%%%%%%%%%%%%%%%%%%%%%%%%%%%%%%%%%%%%%%%%%%%%%%%%%%%%%%%%%%%%%%%%%%%%%%%%%%%%%%%%%%%%%%%%%%%%%%%%%%%%%%%%%%%%%%%%%%%%%%%%%%%%%%%%%%

\section{Discussion and conclusions}\label{disc}

From the analysis of different existing models we conclude that the appearance of ghosts and other particle instabilities, as well as of causality and Lorentz-invariance violation, is not an uncommon feature in Gauss--Bonnet gravity. In particular, we have seen that few of the string-inspired models examined so far are theoretically viable. It is still unclear if there is a nonempty set of initial conditions satisfying all the no-ghost and sub-luminal conditions during the entire evolution of the universe, but it has been established that these constraints do limit, in a decisive way, the parameter space of initial conditions.

Cosmological solutions which are phase-space attractors are not guaranteed to be ghost free. Stability in the phase space $\{H,\dot H,\phi,\dot\phi\}$ against homogeneous perturbations does not imply unitarity at the quantum level. Our analysis covers cosmologically long time intervals, and the possibility of allowing for the existence of unobservable, decaying ghost modes is not realistic within our framework. This fact contributes to ruling out most of the models studied. Some of these are indeed ghost free at any time but perturbations propagate at a speed faster than light. We have also given an example of a model viable from $z\sim 10^4$, but it is motivated only phenomenologically.

Therefore, we can conclude that the presence of ghost modes for the dilaton and moduli fields may be a compelling reason to stabilise them before they can play some role in cosmology. Our results indicate that the non-minimally coupled scalar field of the action \Eq{act} should acquire a precise value through the cosmological dynamics. The alternative to face is to carefully fine tune the initial conditions of the equations of motion, which makes this class of models rather unattractive. However, its study is far from being complete.

The presence of ghosts and superluminal modes in effective Lagrangians obtained by string theory on FRW backgrounds should not surprise the reader. In fact, all these models are based on an effective Lagrangian which was found only assuming a Minkowski background. However, as we have shown in this paper, the same Lagrangians cannot be used \emph{tout court} in cosmology, otherwise unphysical modes may propagate. On the other hand, an effective Lagrangian should not depend in principle on the particular background we choose to study: even the Einstein--Hilbert action is used to describe any kind of gravitational system. It is then possible that cosmology may actually help exclude those actions which give rise to unphysical states. 

To achieve quantum stability and a satisfactory cosmological evolution it is necessary to impose a particular form for the potential. In our case we must either require some nonperturbative mechanism that produces potentials which can heal the FRW (and any other) background from ghosts and superluminal propagation, or we should regard the GB action as an effective one to be corrected by loop contributions. These were found for type II string with fixed moduli in \cite{GZ,FPSS,GrW,Tse86} and in dilaton super and heterotic string in \cite{GS,PZ}. 
The results are summarized in \cite{BB}. We note however that, in general, there are terms which add to higher-order Euler densities and lead to effective theories with spin-2 ghosts even on Minkowski background. As stressed in the introduction, higher-order curvature terms are subdominant at low energies anyway, while at high energies a new and more involved analysis, both of background and perturbed equations from a nonperturbative action, is required.

We have not fully addressed the issue of the acceleration of the universe. Equation \Eq{acceq} gives the necessary information for constraining the evolution of $a$ and $\phi$ so that to reproduce the present inflationary expansion. Although we have performed an analysis of such dynamics in comparison with cosmological data, it is worth further investigation. In particular, we were able to find a ghost-free \emph{and} sub-luminal \emph{and} today-accelerating solution, but not to fully constrain the parameter space to be compatible with experimental bounds from nucleosynthesis until today. We have considerably restricted it, though.

In order to study more realistic (from the string theory point of view) scenarios, one should extend the couplings appearing in Eq.~(\ref{act}) to be dependent on more than one (complex) moduli. This has  recently been considered in \cite{neu06}, and here we have addressed it in a very simplified way. The main conclusion is that the ghost constraints from the tensor modes are unaltered, and those are enough to rule most models out. However, it would be interesting to study more complex scenarios with several moduli fields evolving simultaneously. This immediately leads us to an interesting future direction to explore which is to study how scalar potentials more complicated than Eq.~\Eq{expot} affect the dynamics of the system. From our analysis there is evidence that exponential potentials are hardly able to describe cosmologies which are both free from instabilities and compatible with observations and, in general, one is forced to trade one shortcoming for the other. Considering flux compactification models would combine both the inclusion of more than one scalar field into the analysis, with more involved dynamics parametrised by the flux terms in the scalar potential, and would certainly lead to interesting situations. 

We have chosen a specific form for the functions $f_i$, but these can be more general beyond tree level. For example, one may invoke exponential series from higher-loop corrections, as in \cite{TBF}, or nonperturbative effects. It is clear that the viability of these models depend on the choice for $f_i$, and a number of cases of interest is still to be explored.

Sub-luminal propagation of perturbative modes is one of the conditions which limit the parameter space of the theory. In scenarios where faster-than-light travel is allowed, such as theories defined on non-commutative backgrounds \cite{HI}, we would expect the space of initial conditions to be modified and somehow enlarged. However, the parameter space would be still non-trivial. This is also true when, for some other reason, one abandons the sub-luminal constraints altogether.

Finally, if matter and radiation are non-minimally coupled with $\phi$ the cosmological evolution may change considerably \cite{CTS}.

In this paper we have established, both analytically and numerically, the following very general results:
\begin{itemize}
\item Classical and quantum stability are inequivalent issues which require separate assessments, one not implying the other. Most of the models presented in literature were tested in phase space, but for arbitrary initial conditions they have ghosts or other kinds of particle instabilities, as well as problems related to the propagation speed of particle degrees of freedom.
\item In particular, even the most orthodox models based on string theory (i.e., flat directions like the dilaton or a compactification modulus) are not automatically ghost-free: we are forced to select special initial conditions.
\item It is very difficult, although not impossible, to obtain a theoretically and experimentally viable picture from this class of models. We were able to give only one positive example, with a GB coupling apparently well below any reasonable threshold of experimental detectability. Still, minimally coupled General Relativity is by far the most successful theory at our disposal.
\item Qualitatively, the presence of many noninteracting scalar degrees of freedom does not change the main features of the theory, including the existence of nontrivial selection rules on the parameter space.
\end{itemize}
The presented parameter selection procedure, although remarkably effective, clearly falls short of being fully prescriptive, and there is some evidence that non-minimally coupled GB cosmology is unlikely to be a viable theory for natural values of the parameters and simple choices for the scalar functions. We hope to further clarify this and at least part of the above points in the near future.

%%%%%%%%%%%%%%%%%%%%%%%%%%%%%%%%%%%%%%%%%%%%%%%%%%%%%%%%%%%%%%%%%%%%%%%%%%%%%%%%%%%%%%%%%%%%%%%%%%%%%%%%%%%%%%%%%%%%%%%%%%%%%%%%%%%%%%%%%%%%%%%%%%%%%%%%%%%%%%%%%%%%%%%%%%%%%%%%%%%%%%%%%%%%%%%%%%%%%%%%%%%%%%%%%%%%%%%%%%%%%%%%%%%%%%%%%%%%%%%%

\section*{Acknowledgments}
The authors thank M.\ Hindmarsh, A.R.\ Liddle, P.\ Mukherjee and F.\ Vernizzi for useful discussions.  G.C. (partly), B.d.C. and A.D.F. are supported by PPARC (UK). G.C. is also supported by a Marie Curie Intra-European Fellowship under EC contract MEIF-CT-2006-024523.

%%%%%%%%%%%%%%%%%%%%%%%%%%%%%%%%%%%%%%%%%%%%%%%%%%%%%%%%%%%%%%%%%%%%%%%%%%%%%%%%%%%%%%%%%%%%%%%%%%%%%%%%%%%%%%%%%%%%%%%%%%%%%%%%%%%%%%%%%%%%%%%%%%%%%%%%%%%%%%%%%%%%%%%%%%%%%%%%%%%%%%%%%%%%%%%%%%%%%%%%%%%%%%%%%%%%%%%%%%%%%%%%%%%%%%%%%%%%%%%%

\end{document}